\documentclass[11pt]{article}

\pdfoutput=1 
\usepackage{jheppub}
\usepackage{graphicx,verbatim}
\usepackage{psfrag, times}
\usepackage[T1]{fontenc}
\usepackage{epsf}
\usepackage{mathtools}
\def\be{\begin{equation}}
\def\ee{\end{equation}}
\def\bc{\begin{center}}
\def\ec{\end{center}}

\def\bbF{{\mathbb{F}}}
\def\bbI{{\mathbb{I}}}
\def\bF{{\bf{F}}}

\def\cA{{\mathcal{A}}}
\def\cB{{\mathcal{B}}}
\def\cC{{\mathcal{C}}}
\def\cD{{\mathcal{D}}}

\def\cF{{\mathcal{F}}}
\def\cM{{\mathcal{M}}}
\def\cN{{\mathcal{N}}}

\def\cT{{\mathcal{T}}}
\def\cI{{\mathcal{I}}}
\def\cJ{{\mathcal{J}}}

\def\cK{{\mathcal{K}}}

\def\cO{{\mathcal{O}}}

\def\ga{\gamma}
\def\nn{\nonumber}

\def\Del{{\Delta}}

\def\r2{{\sqrt{2}}}

\def\bz{\bar{z}}

\def\tQ{\tilde{Q}}
\def\tJ{\tilde{J}}
\def\tDel{\tilde{\Del}}
\def\Ga{\Gamma}

\def\bea{\begin{eqnarray}}
\def\eea{\end{eqnarray}}
\def\ru{{\rm{u}}}

\def\cT{{\mathcal{T}}}

\def\Del{\Delta}
\def\ra{{\rm{a}}}
\def\rb{{\rm{b}}}

\def\del{\delta}
\def\ga{\gamma}
\def\tDel{\tilde{\Delta}}

\def\ru{{\rm{u}}}
\def\rv{{\rm{v}}}
\def\ra{{\rm{a}}}
\def\rb{{\rm{b}}}

\def\bc{{\bf{c}}}

\def\rs{{\rm{s}}}
\def\rt{{\rm{t}}}
\def\ru{{\rm{u}}}

\def\veps{\varepsilon}

\def\bJ{\bar{J}}
\def\ow{\omega}
\def\bM{{\bf{M}}}
\def\fg{{\mathfrak{g}}}

\DeclarePairedDelimiter\floor{\lfloor}{\rfloor}

\begin{document}
\title{\Large On Conformal Blocks, Crossing Kernels and Multi-variable Hypergeometric Functions.}
\author[]{Heng-Yu Chen${}^{1}$ and Hideki Kyono${}^{2}$}
\affiliation{$^1$\rm Department of Physics, National Taiwan University, Taipei 10617, Taiwan}
\affiliation{$^2$ \rm Department of Physics, Kyoto University,
Kitashirakawa Oiwake-cho, Kyoto 606-8502, Japan\footnote{
The author belonged there until Mar. 2019.
}}
\affiliation{\tt{ heng.yu.chen@phys.ntu.edu.tw}, \tt{hhh1567@gmail.com}} 
\vspace{2cm}
\abstract
{In this note, we present an alternative representation of the conformal block with external scalars in general spacetime dimensions in terms of a {\it finite} summation over Appell fourth hypergeometric function ${\bF}_4$. We also construct its generalization to the non-local primary exchange operator with continuous spin and its corresponding Mellin representation which are relevant for Lorentzian spacetime. Using these results we apply the Lorentzian inversion formula to compute the so-called crossing kernel in general spacetime dimensions,  the resultant expression can be written as a double infinite summation over certain Kamp\'{e} de F\'{e}riet hypergeometric functions with the correct double trace operator singularity structures. We also include some complementary computations in AdS space, demonstrating the orthogonality of conformal blocks and performing the decompositions.}

\maketitle
\section{Introduction and Summary}
\paragraph{}
Conformal blocks are fundamental building elements for constructing various correlation functions in conformal field theories,
they allow us to disentangle the model-dependent dynamical CFT data from universal kinematical pieces constrained by conformal symmetries. 
Once the operator product expansion (OPE) channel is specified, we can expand conformal correlation functions in terms of conformal blocks, 
and the expansion coefficients are precisely the product of OPE coefficients. 
A similar situation occurs in elementary quantum mechanics, we expand arbitrary wave functions in terms of the eigenfunctions, 
which are often determined by the symmetries and the physical spectra of a given system, and the probability amplitudes describe the underlying dynamics.
These eigenfunctions solve the appropriate Sturm-Liouville equations which ensure their orthogonality, 
moreover the corresponding equations can often be transformed into the defining differential equations of various well-known special functions.
\paragraph{}
It is natural to ask whether conformal blocks can also be systematically studied in the same vein, in other words, whether they arise as the eigenfunctions of suitable Sturm-Liouville type of equations? Are they orthogonal with respect to given integration measure, and can be connected with certain well-studied mathematical functions? 
The answers to these questions are affirmative. Various crucial ingredients have become available starting with the important work of Dolan and Osborn \cite{DO-2003}, \cite{DO-2011}, 
where a second order partial differential equation known as ``quadratic Casimir equation'', determining the four-point global conformal blocks as a function of cross ratios was derived. 
The scaling dimension $\Del$ and spin $J$ of the exchange primary operator in the OPE governs the asymptotic behavior, 
and in even spacetime dimensions, compact solutions can be expressed in terms of Gauss hypergeometric functions ${}_2 F_1$. 
For general spacetime dimensions, conformal blocks were given in terms of an infinite summation over Gegenbauer polynomials \cite{DO-2003} 
or more recently for three dimensions, where the expression is again an infinite summation over ${}_2F_1$s \cite{Hogervorst-2016}.
During these investigations, it was realized that conformal blocks share a close resemblance with a special class of two variable generalization of Jacobi polynomials studied extensively by Koornwinder and collaborators \cite{KoornwinderPapers}. These polynomials are defined through a pair of commuting second and fourth order partial differential equations whose eigenvalues are labeled by two integers\footnote{This operator was also recently considered in \cite{Fortin-2016a}, \cite{Fortin-2019a} in embedding space.}, and their orthogonality can be subsequently defined with known integration measure.
In a related recent development \cite{Schomerus-2016}, \cite{Chen-Qualls}, it was shown that quadratic and quartic conformal Casimir operators can be mapped to linear combinations of two commuting conserved Hamiltonians of the trigonometric $BC_2$ Calogero-Sutherland integrable system via a similarity transformation. 
The conformal blocks can therefore be directly identified with the explicit eigenfunctions of the $BC_2$ Calogero-Sutherland system known as ``Harish-Chandra functions'' \cite{Schomerus1}, which can again be expressed in terms of an infinite summation over hypergeometric functions.
\paragraph{}
Even when the explicit compact form of conformal blocks is available, integration involving them over spacetime coordinates or cross ratios still requires more diligence, as there are a few subtleties.
First, because of the non-compactness of the conformal group, $SO(1, d+1)$ for Euclidean and $SO(2,d)$ for Lorentzian respectively, this means the eigenvalues for conformal Casimir operators are functions of continuous variables $(\Del, J)$. This is similar to the scattering states in quantum mechanics which are generally non-normalizable. 
Second, the naive integration range often extends beyond the radius of OPE convergence for a given channel, additional analytic continuations are needed for the integral to be well-defined.
Besides purely mathematical interests, obtaining an explicit and succinct form of conformal blocks and understanding their general properties also fits well with the conformal bootstrap program (See \cite{Reviews} for reviews). 
Specifically, consider the crossing symmetry equation arises from expanding a single correlation function in terms of the basis for different OPE channels,
in order to extract the OPE coefficients or to relate the spectra in different channels, we need to understand the orthogonality and analytic continuation to perform OPE inversion \cite{CH-2017}, \cite{SSW-2017} or compute the corresponding mixing matrices known as the ``crossing kernel''  or ``6j symbols'' \cite{CrossingKernel-1}, \cite{CrossingKernel-2}, \cite{Liu-2018}, \cite{Karateev-6J}\footnote{See also \cite{Mellin-Bootstrap1}, \cite{Mellin-Crossing}, \cite{Sleight-Taronna CK}, \cite{Sleight-Taronna CK2}  for a closely related computation of crossing kernels in the form of mixing coefficients among Mellin amplitudes.}. In addition, it should be noted that we can consider large spin perturbation, i.e. when the spin of the exchange primary operator becomes large, we can recover the OPE data using the corresponding inversion formula for any twist and all orders from the double discontinuities in cross-ratio plane \cite{Alday}.
\paragraph{}
In this note, we contribute some useful results in several directions towards this systematic study. 
In section \ref{Sec:Block-Hyper}, we provide an alternative expression for conformal block with integer spin exchange in general spacetime dimensions by considering its Mellin-Barnes representation  \cite{Mack:2009}, 
and showing that the integrated result can be written in a {\it finite} summation of Appell's hypergeometric function ${\bf F}_4$ over an integer partition\footnote{It is interesting to note that similar multi-variable generalization of hypergeometric equations were also used to describe the propagation of scalar field in the integrable deformations of $AdS_p \times S^p$ geometries \cite{Lunin-Papers}. It was also brought to our attention after this work is completed, when study conformal correlation functions constrained by conformal Ward identities in the momentum space, expressions in terms of multi-variable hypergeometric functions  can also naturally arise, as discussed in \cite{Coriano-Papers}.}. 
We demonstrate its validity by numerically matching it with known compact expressions in even dimensions and provide several other analytic checks and simplifications which can be useful for conformal bootstrap computations. We then further explicitly construct the conformal block for primary exchange operators with continuous spin by constructing a continuous version of the Mack polynomial or more appropriately named ``Mack function''.
The result can be expressed as linear combination of certain $\fg^{(\rs)}_{\Del, J}(z, \bz)$ functions given in \eqref{Def:fgblock} and its spin shadow transformation\footnote{See also the earlier paper \cite{ChungI} about constructing Lorentzian conformal blocks in a simpler setting.}. 
These results provide us with the necessary ingredients for the computation of the crossing kernel in general spacetime dimensions in Section \ref{Sec:Crossing}, where we applied the Lorentzian inversion formula derived in \cite{CH-2017} (see also \cite{SSW-2017}). We demonstrate the final result can be written in terms of double integrals containing Kamp\'{e} de F\'{e}riet functions, this representation is useful for us to identify the singularities associated with the double trace operators. Finally in section \ref{Sec:AdS}, we provide complementary computations by considering the holographic dual configurations for conformal partial waves in Anti de-Sitter (AdS) space, and demonstrate their orthogonality which can be understood as the orthogonality of AdS harmonic functions. We also consider some simple expansions in terms of these natural bases.
In a few appendices, we provide some useful details about multi-variable hypergeometric functions and computation details.

\section{Conformal Blocks with Integer Spin and Appell's Hypergeometric Functions}\label{Sec:Block-Hyper}
\paragraph{}
In this note we will primarily focus on the four point correlation function of conformal field theories in general spacetime dimensions, 
which contain external scalar primary operators $\cO_{\Del_i, 0}(x_i) = \cO_{i}(x_i)$ with scaling dimension $\Del_i,~i=1, 2,3, 4$:
\be\label{4pt-Scalar}
\left\langle \prod_{i=1}^4 \cO_{i}(x_i)\right\rangle 
= \cT_{\Del_i}^{(\rs)}(x_i)\sum_{\{\cO_{\Del, J}\}}  \lambda_{12\cO} \lambda_{34\cO} \, G_{\Del, J}^{(\rs)}(z, \bz).
\ee
Here we have decomposed the correlation function into summation over the $s$-channel exchange primary operator $\{\cO_{\Del, J}(x)\}$
and $ \lambda_{12\cO} \lambda_{34\cO}$ are the product of OPE coefficients. 
Here we can further factor out the overall kinematic factor depending only on the external scaling dimensions $\{\Del_i\}$:
\be\label{Kfactor-s}
\cT^{(\rs)}_{\Del_i} (x_i) = \frac{1}{(x_{12}^2)^{\frac{\Del_{12}^+}{2}}(x_{34}^2)^{\frac{\Del_{34}^+}{2}}} \left(\frac{x_{14}^2}{x_{24}^2}\right)^{\ra^{(\rs)}} \left(\frac{x_{14}^2}{x_{13}^2}\right)^{\rb^{(\rs)}}, \quad \ra^{(\rs)} = \frac{\Del_{21}^-}{2}, ~~ \rb^{(\rs)} = \frac{\Del_{34}^-}{2},
\ee
and introduce the notation $\Del_{ij}^{\pm} = (\Del_i\pm \Del_j)$. 
The remaining function $G_{\Del, J}^{(\rs)}(z, \bz)$ is called the s-channel conformal block which depends only on a pair of conformally invariant cross ratios
$(\ru, \rv)$ or $(z,\bz)$:
\be\label{Def:CR1}
\ru=z\bz  = \frac{x_{12}^2 x_{34}^2}{x_{13}^2 x_{24}^2}, ~~ \rv=(1-z)(1-\bz)  = \frac{x_{14}^2 x_{23}^2}{x_{13}^2 x_{24}^2}, \quad x_{ij}^2 = (x_i-x_j)^2.
\ee
In Euclidean spacetime ${\mathbb R}^d$, $x_{ij}^2 \ge 0$ which implies $\ru, \rv \ge 0$ or the cross ratios $(z, \bz)$ are complex conjugate to each other $\bz = z^*$. 
While in Lorentzian spacetime ${\mathbb M}^{1, d-1}$, $x_{ij}^2$ can become negative instead, $(z, \bz)$ are treated as independent complex variables in general,
this distinction becomes important when we consider integration over the conformal blocks. 
\paragraph{}
It is well known that the conformal block $G_{\Del, J}^{(\rs)}(z, \bz)$ satisfies a so-called quadratic Casimir equation, which is a second order partial differential equation in $(z, \bz)$ 
and is parameterized by $(\ra^{(\rs)}, \rb^{(\rs)})$\cite{DO-2011}:
\be\label{Casimir2-eqn}
\left(D_z(\ra^{(\rs)},\rb^{(\rs)})+ D_{\bz}(\ra^{(\rs)},\rb^{(\rs)})+2\veps\frac{z\bz}{z-\bz}\left((1-z)\frac{\partial}{\partial z}- (1-\bz)\frac{\partial}{\partial \bz}\right)\right)G_{\Del, J}(z,\bz) = C_2(\Del, J)G_{\Del, J}(z,\bz),
\ee
where:
\be\label{Dab}
D_z(\ra, \rb) = z^2(1-z)\frac{d^2}{dz^2} - (1+\ra+\rb)z^2\frac{d}{dz}-\ra\rb z, \quad h=\frac{d}{2},~~\veps = h-1,
\ee
and the eigenvalue $C_2(\Del, J)$:
\be\label{Def:C2}
C_{2}(\Del, J) = \Del(\Del-d)+J(J+d-2),
\ee
depends on the scaling dimension $\Del$ and the spin $J$ of the exchanged operator $\cO_{\Del, J}(x)$.
It is interesting to note that treating $(\Del, J)$ as parameters, the eigenvalue $C_{2}(\Del, J)$ has following three independent ${\mathbb{Z}}_2$ symmetry transformations:
\be\label{8fold-sym}
{\text{Shadow}}:\Del \Leftrightarrow d-\Del, ~~~{\text{Spin Shadow}}: J\Leftrightarrow 2-d-J,~~~{\text{Light Ray}}:\Del \Leftrightarrow 1-J,
\ee
these imply that \eqref{Casimir2-eqn} has eight-fold degeneracies of independent eigenfunctions, only distinguished by their different asymptotic behaviors.
The first one in \eqref{8fold-sym} corresponds to the well-known shadow transformation introduced in \cite{DO-2011} which maps a local primary $\cO_{\Del, J}(x)$
to a non-local operator $\tilde{\cO}_{d-\Del, J}(x)$ with shadow scaling dimension $d-\Del$ via an integral transformation.
The remaining two transformations in \eqref{8fold-sym} generally map $J$ to negative or even continuous values, 
as explained in details in \cite{LightRay} (The close connection of these transformation with the Weyl group of $BC_2$ group was also detailed in \cite{Schomerus1}), these values can only be physical in Lorentzian instead of Euclidean spacetime. A primary operator $\cO_{\Del, J}(x)$ now transforms under $SO(2, d)$ instead of $SO(1,d+1)$, it is labeled by the two continuous Cartan numbers $(\Del, J)$ of $SO(1,1)_\Del \times SO(1,1)_J \in SO(2, d)$, and they can be mixed under the Weyl group of $SO(2, d)$. It is useful to introduce here the spectral parameterization for Lorentzian spacetime:
\be\label{Def:Spec2}
\Del = h+i\nu, \quad J = -\veps +i \ell,
\ee
such that the eigenvalue \eqref{Def:C2} can be rewritten as:
\be\label{Def:C2-1}
C_2(\nu, l) = (h^2 +\nu^2) + (\veps^2 + \ell^2).
\ee
The transformations in \eqref{8fold-sym1} can be succinctly recast in terms of spectral parameters $(\nu, l)$:
\be\label{8fold-sym1}
{\text{Shadow}}: \nu \Leftrightarrow -\nu, ~~~{\text{Spin Shadow}}: \ell \Leftrightarrow -\ell,~~~{\text{Light Ray}}: \nu \Leftrightarrow -\ell,
\ee
which clearly leave \eqref{Def:C2-1} invariant.
We will return to these parameterizations when we consider the conformal block with continuous spin and the inversion formula of \cite{CH-2017}.
\paragraph{}
In this note, we will ultimately be interested in constructing so-called ``crossing kernel'' for general $d$-dimensional conformal field theories, which will be introduced momentarily. 
{
To begin with, we first construct an orthogonal basis in ${\mathbb{R}}^d$ which will be called conformal partial wave (CPW) \cite{DO-2011}:
\be\label{Def:psi}
\Psi^{(\rs)}_{\nu,J}(x_i)=\frac{1}{\pi^h}
\int_{\mathbb{R}^d} d^dx_0 
\langle
\cO_{\Delta_1}(x_1)
\cO_{\Delta_2}(x_2)
\cO_{h+i\nu,J}^{\mu_1...\mu_J}(x_0)
\rangle
\langle
\tilde{\cO}_{h-i\nu,J}^{\mu_1...\mu_J}(x_0)
\cO_{\Delta_3}(x_3)
\cO_{\Delta_4}(x_4)
\rangle.
\ee
Here the two copies of three point functions contain $\cO_{h+i\nu, J}^{\mu_1 \dots \mu_J }(x_0)$ and its shadow $\tilde{\cO}_{h-i\nu, J}^{\mu_1 \dots \mu_J }(x_0)$, 
they are related through following integral transformation:
\be\label{Def:Shadow}
\tilde{\cO}_{h-i\nu, \mu_1 \dots, \mu_J} (x)= \frac{1}{\pi^h}\frac{\Ga(h-i\nu+J)}{(h+i\nu-1)_J \Ga(i\nu)}\int d^d x' \frac{\cI_{\mu_1, \dots, \mu_J; \nu_1, \dots, \nu_J}(x-x')}{((x-x')^2)^{h-i\nu}}\cO_{h+i\nu, \nu_1, \dots, \nu_J}(x'),
\ee
where $\cI_{\mu_1, \dots, \mu_J; \nu_1, \dots, \nu_J}(y)$ is the inversion tensor of symmetric traceless tensors.
The functional forms of these three point functions are fixed kinematically through conformal symmetry\footnote{In this paper we follow the normalization convention of conformal partial wave $\Psi_{\nu, J}^{(
\rs)}(x_i)$ explicitly constructed in \cite{DO-2011} by fusing two copies of three point functions together, this also fixes our expansion coefficients hence the normalization of conformal blocks.}.
Notice that the CPW is also an eigenfunction of the conformal Casimir equation and it can be expanded 
in terms of a conformal block and its shadow as follows:
\be\label{Def:Psinu1}
\Psi_{\nu, J}^{(\rs)}(x_i) 
= \cT^{(\rs)}_{\Del_i} (x_i) \sum_{\sigma_s =\pm} \bc_{h+i\sigma_s \nu, J}^{(\rs)} G_{h+i\sigma_s \nu, J}^{(\rs)}(z,\bz)
\ee
}
where the expansion coefficients are given by
\be\label{Def:cnu-coeff}
\bc_{h+i\sigma_s\nu, J}^{(\rs)} =\frac{1}{2^J c_J} \frac{(h-i\sigma_s\nu-1)_J \Gamma(-i\sigma_s\nu)}{ \Gamma(2\ow_{\sigma_s})} \frac{ \Ga\left(\ow_{\sigma_s}\pm \ra^{(\rs)}\right)  \Ga\left(\ow_{\sigma_s}\pm\rb^{(\rs)}\right)}{ \prod_{i=1}^4 \Ga\left(\gamma^{(\rs)}_i\right)},
\ee  
with  $c_J = \frac{(\veps)_J}{(2\veps)_J}$ and 
$(x)_J =\frac{\Ga(x+J)}{\Ga(x)}$ being the Pochhammer symbol. 
In the above, we have also introduced the notation $\Gamma(x\pm a) = \Ga(x+a)\Ga(x-a)$.
We have also defined the following combinations of scaling dimensions and spins:
\be\label{Def:tau-s}
\ow_{\sigma_s } = \frac{h+i\sigma_s\nu+J}{2}, \quad 
\tau_{\sigma_s } = \frac{h+i\sigma_s\nu-J}{2}, ~~ \sigma_s =\pm.
\ee
while $\{\gamma_i^{(\rs)}\}$ are given in \eqref{gamma0i-s}.
The integer spin-$J$ operators $\cO_{h+i\nu, J}(x_0)$ and $\tilde{\cO}_{h-i\nu, J}(x_0)$  carry the complex scaling dimensions, 
we can recover the conformal block with physical real dimensions by integrating the conformal partial wave \eqref{Def:Psinu1} over the $\nu$-plane together with the spectral function for $\nu$, see for example \cite{CKK:2017}.
\paragraph{} 
It is well-known that the conformal partial wave \eqref{Def:Psinu1} also enjoys the Mellin representation \cite{DO-2011}, \cite{Mack:2009}:
\be\label{Psi-s-Mellin1}
{\Psi}_{\nu, J}^{(\rs)}(x_i)  
=\cT_{\Del_i}^{(\rs)}(x_i) \int^{i\infty}_{-i\infty} \frac{ds}{(4\pi i)}  \int^{i\infty}_{-i\infty} \frac{dt}{(4\pi i)} \ru^{\frac{s}{2}} \rv^{\frac{t}{2}}  \rho^{(\rs)}_{\Del_i} (s, t) \cM^{(\rs)}_{\nu, J}(s, t)\,.  
\ee
Here the Mellin space integration measure is given by:
\be\label{rho-s}
\rho^{(\rs)}_{\Del_i} (s, t) = \prod_{i< j} {\Ga\left(\del_{ij}^{(\rs)}\right)},
\ee 
where in principle we have three Mandelstam-like variables $(s, t, u)$ satisfying $s+t+u=\sum_{i=1}^4\Del_i$, such that we can substitute $u$ away and express $\{\delta_{ij}^{(\rs)}\}$ only in terms of $(s, t)$:
\be\label{deltaij}
\del_{12}^{(\rs)} = \frac{\Del_{12}^+ -s}{2},~\del_{34}^{(\rs)} = \frac{\Del_{34}^+ -s}{2},~
\del_{13}^{(\rs)} = \frac{s+t}{2} +\rb^{(s)},~\del_{24}^{(\rs)} = \frac{s+t}{2}+\ra^{(s)},~
\del_{14}^{(\rs)} = -\frac{t}{2} -\ra^{(\rs)}-\rb^{(\rs)},~\del_{23}^{(\rs)} = -\frac{t}{2},
\ee 
such that $\sum_{j \neq i} \delta_{ij}^{(\rs)} = \Del_i$.  
Notice the number of independent Mellin parameters is exactly the same as the independent cross ratios, we can regard the Mellin representation as
trading $(\ru, \rv)$ with $(s, t)$.
While the remaining integrand:
\be\label{PMellinAmp-s}
\cM^{(\rs)}_{\nu, J}(s, t) = \frac{1}{\prod_{i=1}^4 \Ga\left(\gamma^{(\rs)}_i\right)} \frac{\Ga\left(\tau_{\pm}-\frac{s}{2}\right)}{\Ga\left(\del_{12}^{(\rs)}\right) \Ga\left(\del_{34}^{(\rs)}\right)}{P}_{\nu, J}^{(\rs)} (s, t),
\ee
is the corresponding $s$-channel partial Mellin amplitude.
The polynomial ${P}_{\nu, J}^{(\rs)} (s, t)$ originally due to Mack \cite{Mack:2009} is given by: 
\be\label{Mack-poly1}
{P}_{\nu, J}^{(\rs)} (s, t) ={\widetilde{\sum_{r,k}}} \left(\tau_{\pm }-\frac{s}{2}\right)_r  \prod_{(ij)}{\left(\delta_{ij}^{(\rs)}\right)_{k_{ij}}}
\prod_{i=1}^4 \frac{\Ga\left(\gamma^{(\rs)}_{i}\right)}{\Ga\left(\gamma^{(\rs)}_{i}-J+r+\sum_{j}k_{ji}\right)},
\ee
where the abridged double summations:
\be\label{Def:tilde-summation}
{\widetilde{\sum_{r,k}}} \dots =\frac{J!}{2^J } \sum_{r=0}^{\floor*{\frac{J}{2}}} (-1)^r \frac{(J+\veps)_{-r}}{r! (J-2r)!} \sum_{\sum k_{ij} = J-2r} (-1)^{k_{13}+k_{24}} \frac{(J-2r)!}{\prod_{(ij)}k_{ij}!} \dots
\ee
is over $r$ which comes from the expansion coefficients of Gegenbauer polynomial arising from tensor contraction of three point functions, $\floor*{\frac{J}{2}}$ denotes integer part of $\frac{J}{2}$
and the four-fold integer partition $\{k_{ij}\}$ with $(ij) = \{13, 24, 14, 23\}$ such that $\sum_{(ij)}k_{ij} = J-2r$. 
The remaining parameters $\{\ga_{i}^{(\rs)}\}$ arising from the integration via Symanzik formula are given by:
\be\label{gamma0i-s}
\ga_{1}^{(\rs)} = \ow_{+}-\ra^{(\rs)},~ \ga_{2}^{(\rs)} = \ow_{+}+\ra^{(\rs)},~
\ga_{3}^{(\rs)} = \ow_{-}+\rb^{(\rs)},~ \ga_{4}^{(\rs)} = \ow_{-}-\rb^{(\rs)}.
\ee
Despite the rather complicated looking expression for ${P}_{\nu, J}^{(\rs)} (s, t)$ in \eqref{Mack-poly1}, the most important analytic information we need about it is that it does not contain any additional singularities in $(s,t)$ variables, but only zeroes which cancel some of the poles in the $(s, t)$-contour integrations. It is also useful to note that Mack polynomial \eqref{Mack-poly1} enjoys a somewhat non-obvious but important reflection symmetry ${P}^{(\rs)}_{\nu, J}(s, t) = {P}^{(\rs)}_{-\nu, J}(s, t)$ \cite{ConfRegge}, which is needed for consistently extracting the conformal block from integration over $\nu$-plane.
\paragraph{}
Using the Mellin representation of  $\Psi^{(\rs)}_{\nu, J}(x_i)$ \eqref{Psi-s-Mellin1}, we can readily integrate it into a summation over regularized Gauss Hypergeometric functions ${}_2 \tilde{F}_1$:
\bea\label{Psi-s-full}
\Psi^{(\rs)}_{\nu, J}(x_i) 
&=&\cT^{(\rs)}_{\Del_i}(x_i)
\sum_{\sigma_s =\pm}
\widetilde{\sum_{r, k}}
\prod_{i=1}^4 \frac{1}{\Ga\left(\gamma^{(\rs)}_{i}-J+r+\sum_{j}k_{ji}\right)}
\frac{\ru^{\tau_{\sigma_s }+r} }{\rv^{\frac{\ra^{(\rs)}+\rb^{(\rs)}-k_{14}-k_{23}}{2}}}
\sum_{n_s = 0}^{\infty}  \frac{(-1)^{n_s}\ru^{n_s}}{n_s!} \Ga(-i\sigma_s \nu-n_s)\nn\\
&\times&
\left[ \rv^{\frac{\varpi_{14}^{(\rs)}}{2}} {}_2\tilde{\bf{F}}_1 
 \left[
\begin{matrix}
~\kappa_{\sigma_s}^{1(\rs)}+n_s+r,~\kappa_{\sigma_s}^{4(\rs)}+n_s+r\\
~1+ \varpi^{(\rs)}_{14}
\end{matrix}; \rv
\right] 
+  
\rv^{\frac{\varpi_{23}^{(\rs)}}{2}} {}_2\tilde{\bf{F}}_1 
 \left[
\begin{matrix}
~\kappa_{\sigma_s}^{2(\rs)}+n_s+r,~\kappa_{\sigma_s}^{3(\rs)}+n_s+r\\
~1+ \varpi^{(\rs)}_{23}
\end{matrix}; \rv
\right]\right]\nn\\
&=&\cT^{(\rs)}_{\Del_i}(x_i)
\sum_{\sigma_s =\pm}
\widetilde{\sum_{r, k}}
\prod_{i=1}^4 \frac{1}{\Ga\left(\gamma^{(\rs)}_{i}-J+r+\sum_{j}k_{ji}\right)}
 \frac{ \ru^{\tau_{\sigma_s \nu}+r} }{ \rv^{\frac{\ra^{(\rs)}+\rb^{(\rs)}-k_{14}-k_{23}}{2}}}
\nn\\
&\times&
{\left[ \rv^{\frac{\varpi_{14}^{(\rs)}}{2}} \tilde{\bf{F}}_4 
 \left[
\begin{matrix}
~\kappa_{\sigma_s}^{1(\rs)}+r,~\kappa_{\sigma_s}^{4(\rs)}+r\\
~1+i\sigma_s\nu,~1+\varpi^{(\rs)}_{14},
\end{matrix}; \ru, \rv
\right]
+\rv^{\frac{\varpi_{23}^{(\rs)}}{2}} \tilde{\bf{F}}_4 
 \left[
\begin{matrix}
~\kappa_{\sigma_s}^{2(\rs)}+r,~\kappa_{\sigma_s}^{3(\rs)}+r\\
~1+i\sigma_s\nu,~1+\varpi^{(\rs)}_{23},
\end{matrix}; \ru, \rv
\right]
\right]},
\eea
where we have defined the following combinations:
\bea\label{Def:kappa-s}
&&\kappa_{\sigma_s}^{1 (\rs)} = \tau_{\sigma_s}-\ra^{(\rs)}+k_{13}+k_{14},~ \kappa_{\sigma_s}^{2 (\rs)} =  \tau_{\sigma_s} +\ra^{(\rs)}+k_{23}+k_{24},\nn\\
&&\kappa_{\sigma_s}^{3 (\rs)}=   \tau_{\sigma_s}+\rb^{(\rs)}+k_{13}+k_{23} ,~ \kappa_{\sigma_s}^{4 (\rs)} = \tau_{\sigma_s} -\rb^{(\rs)}+k_{14}+k_{24},
\\
&&\varpi_{14}^{(\rs)} = (\kappa_{\sigma_s}^{1 (\rs)}+r)+(\kappa_{\sigma_s}^{4 (\rs)}+r)-(h+i\sigma_s\nu)= -(\ra^{(\rs)}+\rb^{(\rs)}+k_{23}-k_{14}),\label{vw14}\\ 
&&\varpi_{23}^{(\rs)} =(\kappa_{\sigma_s}^{2 (\rs)}+r)+(\kappa_{\sigma_s}^{3 (\rs)}+r)-(h+i\sigma_s\nu) = \ra^{(\rs)}+\rb^{(\rs)}+k_{23}-k_{14},\label{vw23}
\eea
and the function:
\be\label{bfF21}
{}_2\tilde{\bf{F}}_1 
 \left[
\begin{matrix}
~a,~b\\
~c
\end{matrix}; x
\right] =\frac{\pi}{\sin\pi c} \Ga(a)\Ga(b) {}_2\tilde{F}_1 
 \left[
\begin{matrix}
~a,~b\\
~c
\end{matrix}; x
\right] =\frac{\pi}{\sin\pi c}\int^{+i\infty}_{-i\infty} \frac{dt}{2\pi i} \frac{\Ga(-t)\Ga(a+t)\Ga(b+t)}{\Ga(c+t)} (-x)^{t}.
\ee
Moreover in the second equality of \eqref{Psi-s-full}, 
we notice that for fixed $\{r, k_{ij}\}$, 
the infinite summation over $n_s$ labeling the conformal descendants can also be performed, and the final result is expressed in terms of  Appell's fourth hypergeometric function ${\bf{F}}_4$.
Appell's hypergeometric functions are two parameter generalization of Gauss's hypergeometric function. For a nice introduction, please see \cite{Schlosser}, and we will follow the series definition there for ${\bf{F}}_4$ as given in \eqref{Def:F4} and summarized the relevant details in Appendix \ref{Appendix:Hyper}. Here we have also defined the following function:
\bea
&& \tilde{\bf{F}}_{4}
 \left[
\begin{matrix}
~a_1,~a_2\\
~c_1,~c_2
\end{matrix}; x, y
\right] 
=
\Ga(a_1)\Ga(a_2)\Ga(1-c_1)\Ga(1-c_2)
{\bf{F}}_4   \left[
\begin{matrix}
~a_1,~a_2\\
~c_1,~c_2
\end{matrix}; x, y
\right]\nn\\
&&=\frac{\pi^2}{\sin\pi c_1 \sin\pi c_2} \int^{i\infty}_{-i\infty}\frac{dt}{2\pi i} \int^{i\infty}_{-i\infty}\frac{ds}{2\pi i} \frac{\Ga(a_1+s+t)\Ga(a_2+s+t)\Ga(-s)\Ga(-t) }{\Ga(c_1+s)\Ga(c_2+t)} (-x)^s(-y)^{t}\,,\nn\\
\label{Def:tF4} 
\eea
to absorb additional $\Ga$-functions. In the last equality above, we introduced a Mellin-Barnes representations for $\tilde{\bF}_4$ and the integration contours close in the right half planes to enclose the poles where $s, t \in {\mathbb{N}}_{\ge 0}$.
\paragraph{}
Naively each individual $\tilde{\bF}_4$ in \eqref{Psi-s-full} is strictly convergent when $|\ru|^{\frac{1}{2}}+|\rv|^{\frac{1}{2}} <1$, which is impossible in Euclidean signature $\bz = z^*$,
however $ \Psi^{(\rs)}_{\nu, J}(x_i) $  and the conformal blocks $G_{h\pm i\nu, J}^{(\rs)}(z, \bz)$ are convergent near $(\ru, \rv) = (0,1)$ which is allowed in both Euclidean and Lorentzian spacetimes. 
To see this, we can use a well-known hypergeometric function identity \eqref{hyper-id1}
to rewrite \eqref{Psi-s-full} as follows
\bea\label{Psi-s-full-2}
\Psi^{(\rs)}_{\nu, J}(x_i) 
&=&\cT^{(\rs)}_{\Del_i}(x_i)
%%%%%%%%%%%%%%%%
\sum_{\sigma_s =\pm} \widetilde{\sum_{r, k}} \prod_{i=1}^4 \frac{1}{\Ga\left(\gamma^{(\rs)}_{i}-J+r+\sum_{j}k_{ji}\right)}  \ru^{\tau_{\sigma_s }+r}  \rv^{k_{23}}
\nn\\
&\times& \sum_{n_s = 0}^{\infty}  \frac{(-1)^{n_s} \ru^{n_s} }{n_s!} \Ga(-i\sigma_s \nu-n_s)\prod_{i=1}^4\Ga\left(\kappa_{\sigma_s}^{i(\rs)}+n_s+r\right)
{}_2\tilde{F}_1 \left[
\begin{matrix}
~\kappa_{\sigma_s}^{2(\rs)}+n_s+r, ~\kappa_{\sigma_s}^{3(\rs)}+n_s+r \\
~h+i\sigma_s\nu+2n_s
\end{matrix}; 1-\rv \right],\nn\\
\eea
which is clearly convergent near $(\ru, \rv) = (0, 1)$. 
This form \eqref{Psi-s-full-2} is also useful for us to analytically continue to other convergent series for other values $(\ru, \rv)$, e.g. near $(\ru, \rv) = (0, \infty)$,
we can use another hypergeometric identity \eqref{hyper-id2}
to further rewrite \eqref{Psi-s-full-2} into:
\bea\label{Psi-s-full-3}
\Psi^{(\rs)}_{\nu, J}(x_i) 
&=&\cT^{(\rs)}_{\Del_i}(x_i)
%%%%%%%%%%%%%%%%
\sum_{\sigma_s =\pm} \widetilde{\sum_{r, k}}\prod_{i=1}^4 \frac{1}{\Ga\left(\gamma^{(\rs)}_{i}-J+r+\sum_{j}k_{ji}\right)}  \left(\frac{\ru}{\rv}\right)^{\tau_{\sigma_s }+r} 
\nn\\
&\times&
\left[\frac{1}{ \rv^{\rb^{(\rs)}+k_{13}}} \tilde{\bf{F}}_4 
 \left[
\begin{matrix}
~\kappa_{\sigma_s}^{1(\rs)}+r,~\kappa_{\sigma_s}^{3(\rs)}+r\\
~1+i\sigma_s\nu,~1+\kappa_{\sigma_s}^{3(\rs)}-\kappa_{\sigma_s}^{2(\rs)},
\end{matrix}; \frac{\ru}{\rv}, \frac{1}{\rv}
\right]
+\frac{1}{ \rv^{\ra^{(\rs)}+k_{24}}} \tilde{\bf{F}}_4 
 \left[
\begin{matrix}
~\kappa_{\sigma_s}^{2(\rs)}+r,~\kappa_{\sigma_s}^{4(\rs)}+r\\
~1+i\sigma_s\nu,~1+\kappa_{\sigma_s}^{2(\rs)}-\kappa_{\sigma_s}^{3(\rs)},
\end{matrix}; \frac{\ru}{\rv}, \frac{1}{\rv}
\right]\right]
\,\nn\\
\eea
which are clearly convergent near $(\ru, \rv) = (0, \infty)$.
The expressions in \eqref{Psi-s-full-2} and \eqref{Psi-s-full-3} are also useful for computing the non-trivial monodromy around $\bz =1$ and $\bz = -\infty$ when we apply the Lorentzian inversion formula \cite{CH-2017}.
\paragraph{}
By integrating over the spectral parameter $\nu$, we can obtain the expression for the d-dimensional conformal blocks  $G_{\Del, J}^{(\rs)}(z, \bz)$ with integer spin $J$ \footnote{
Note that for the scalar $J=0$ exchange, it was also noted in \cite{Fortin-2016} that the conformal block can be written in terms of Appell's ${\bf{F}}_4$ functions, our result obtained from a different approach generalizes it to arbitrary integer $J$.}:
\bea\label{Gblock-s}
&&G^{(\rs)}_{\Del, J}(z, \bz) = \frac{1}{\bc_{\Del, J}^{(\rs)}}\widetilde{\sum_{r, k}} \prod_{i=1}^4\frac{1}{\Gamma\left(\gamma_{i}^{(\rs)}(\tau)-J+r+\sum_j k_{ij}\right)}
\ru^{\frac{\Del-J}{2}+r} \rv^{k_{23}}\nn\\
&&\times \sum_{n_s=0}^\infty \frac{(-1)^{n_s}\ru^{n_s}}{n_s!} \Ga(h-\Del-n_s)  \prod_{i=1}^4\Ga\left(\kappa_+^{i(\rs)}(\tau)+r+n_s\right) {}_2\tilde{F}_1 \left[
\begin{matrix}
~\kappa_+^{2(\rs)}(\tau)+r+n_s, ~\kappa_+^{3(\rs)}(\tau)+r+n_s\\
~\Del+2n_s
\end{matrix}; 1-\rv \right]\nn\\
&&=\frac{1}{\bc^{(\rs)}_{\Del, J}}\widetilde{\sum_{r, k}}\prod_{i=1}^4 \frac{ 1}{\Ga\left(\gamma^{(\rs)}_{i}(\tau)-J+r+\sum_{j}k_{ji}\right)} \frac{ \ru^{\frac{\Del-J}{2}+r} }{ \rv^{\frac{\ra^{(\rs)}+\rb^{(\rs)}-k_{14}-k_{23}}{2}}}
\nn\\
&&
\times
{\left[ \rv^{\frac{\varpi_{14}^{(\rs)}}{2}} \tilde{\bf{F}}_4 
 \left[
\begin{matrix}
~\kappa_+^{1(\rs)}(\tau)+r,~\kappa_+^{4(\rs)}(\tau)+r\\
~1-h+\Del,~1+\varpi_{14}^{(\rs)}
\end{matrix}; \ru, \rv
\right]
+\rv^{\frac{\varpi_{23}^{(\rs)}}{2}} \tilde{\bf{F}}_4 
 \left[
\begin{matrix}
~\kappa_+^{2(\rs)}(\tau)+r,~\kappa_+^{3(\rs)}(\tau)+r\\
~1-h+\Del,~1+\varpi_{23}^{(\rs)}
\end{matrix}; \ru, \rv
\right]
\right]},\nn\\
\eea
where $\gamma^{(\rs)}_i(\tau)$ and $\kappa_+^{i(\rs)}(\tau)$ denote we have set $\tau_{+}=\tau=\frac{\Del-J}{2}$ in $\gamma^{(\rs)}_i$ and $\kappa^{i(\rs)}_{+ }$ respectively.
We can check that \eqref{Gblock-s} satisfies the shadow identity:
\be\label{Shadow-Id}
G_{\Del, J}^{(\rs)}(z,\bz)\mid_{\Del_i \to \bar{\Del}_i =d -\Del_i} = ((1-z)(1-\bz))^{\ra^{(\rs)}+\rb^{(\rs)}} G_{\Del, J}^{(\rs)}(z,\bz).
\ee
In the $J=0$, i.e. scalar exchange limit, using the same hypergeometric identity for obtaining \eqref{Psi-s-full-3}, we can rewrite \eqref{Gblock-s} into:
\bea\label{Gblock-s-scalar}
G^{(\rs)}_{\Del, 0}(z, \bz) &=&\ru^{\frac{\Del}{2}}\sum_{n =0}^{\infty}\frac{\ru^{n}}{n !}\frac{\left(\frac{\Del}{2}-\ra^{(\rs)}\right)_n\left(\frac{\Del}{2}-\rb^{(\rs)}\right)_n}{(\Del-h+1)_n}
\sum_{m=0}^{\infty}\frac{\left(\frac{\Del}{2}+\ra^{(\rs)}\right)_{m+n} \left(\frac{\Del}{2}+\rb^{(\rs)}\right)_{m+n}}{m! (\Del)_{2n+m}} (1-\rv)^m\nn\\
&=&\frac{\ru^{\frac{\Del}{2}}}{\bc_{\Del, 0}^{(\rs)}} \left[ 
 \tilde{\bf{F}}_4 
 \left[
\begin{matrix}
~\frac{\Del}{2}+\ra^{(\rs)},~\frac{\Del}{2}+\rb^{(\rs)}\\
~1-h+\Del,~1+\ra^{(\rs)}+\rb^{(\rs)}
\end{matrix}; \ru, \rv
\right]
+
\frac{1}{\rv^{\ra^{(\rs)}+\rb^{(\rs)}}} \tilde{\bf{F}}_4 
 \left[
\begin{matrix}
~\frac{\Del}{2}-\ra^{(\rs)},~\frac{\Del}{2}-\rb^{(\rs)}\\
~1-h+\Del,~1-\ra^{(\rs)}-\rb^{(\rs)}
\end{matrix}; \ru, \rv
\right]
\right],\nn\\
\eea
which precisely coincides with the general $d$-dimensional scalar conformal block given in \cite{DO-2011}.
Numerically we have also checked that \eqref{Gblock-s} has the following desired small $(z, \bz)$ asymptotic behaviors given in \cite{DO-2011}\footnote{Unless otherwise stated, here we follow the normalization of conformal block given in \cite{DO-2011}.}:
\bea
&&G_{\Del, J}^{(\rs)} (z, \bz) \to  \frac{(\veps)_J}{(2\veps)_J} \bz^{\frac{\Del-J}{2}} {}_2 F_1\left[\begin{matrix}~\frac{\Del+J}{2}+\ra^{(\rs)}, ~\frac{\Del+J}{2}+\rb^{(\rs)}\\ \Del+J \end{matrix}; z\right] 
+\cO\left(\bz^{\frac{\Del-J}{2}+1}\right), ~~ 1\gg z \gg \bz \gg 0,\nn\\
\label{Gblock-limit1}\\
&& G_{\Del, J}^{(\rs)} (z, \bz) \to \frac{J!}{(2\veps)_J} (z\bz)^{\frac{\Del}{2}} C_J^{(\veps)}\left(\sigma\right)+\cO\left((z\bz)^{\frac{\Del}{2}+1}\right), ~~z\bz \to 0,~~\sigma = \frac{z+\bz}{2\sqrt{z\bz}}~{\rm fixed}
\label{Gblock-limit2}
\eea
and coincide with the finite sum of hypergeometric functions in terms of cross ratio $(z, \bz)$ when $d=2, 4$,
as given in \cite{DO-2011}. 
%%%%%%%%%%%%%%%%%%%%%%%%%
\begin{figure}[h]
\centering
\begin{tabular}{cc}
\includegraphics[scale=.33]{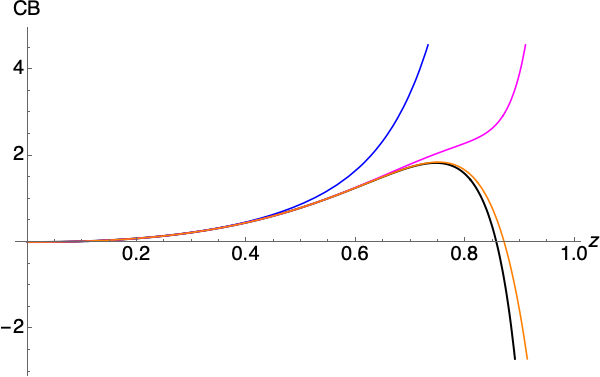}~~ &~~
\includegraphics[scale=.33]{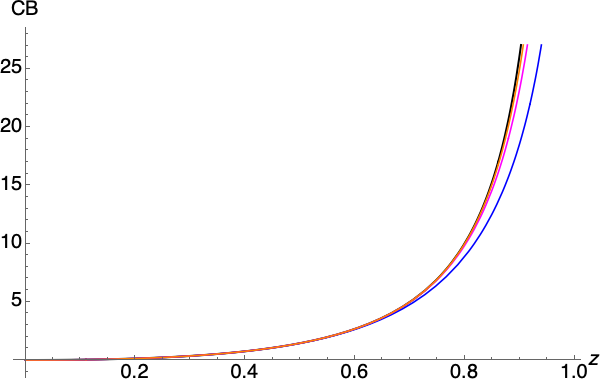} 
\vspace*{0.2cm}\\
{\footnotesize (a) $d=2$ and $(\Del, J) = (2.1, 3)$} &
{\footnotesize (b) $d=4$ and $(\Del, J) = (2.1, 2)$} 
\end{tabular}
\caption{{Numerical match of \eqref{Gblock-s} with known 2 and 4-dimensional conformal blocks in \cite{DO-2011} for each $(\Delta,J)$ and  $\ra^{(\rs)} =\rb^{(\rs)}=0$. We have also set $z =\bz$ in the horizontal axis.
The black curves correspond to the conformal blocks in \cite{DO-2011} and
the blue, pink and orange curves correspond to the conformal block expansions given in \eqref{Gblock-s}
which are truncated at $n_s=2$, $4$ and $6$ respectively. We can see that as we increase the number of terms in the summation for Appell's $\bF_4$ function, our result quickly approaches the known existing results given by the black curves.}
}
\label{Fig:IntegerCB}
\end{figure}
%%%%%%%%%%%%%%%%%%%%%%%%%
By comparing \eqref{Gblock-s-scalar} with \eqref{Gblock-s}, we can also regard conformal block with integer spin $J$ as the partition summation of scalar conformal blocks with shifted external scaling dimensions:
\be
G^{(\rs)}_{\Del, J}(z, \bz) = \frac{\bc_{\Del, 0}^{(\rs)}}{\bc_{\Del, J}^{(\rs)}}\widetilde{\sum_{r, k}} \prod_{i=1}^4\frac{\Ga\left(\gamma_{i}^{(\rs)}(\tau)\right)}{\Gamma\left(\gamma_{i}^{(\rs)}(\tau)-J+r+\sum_j k_{ij}\right)}\frac{1}{\ru^{\frac{J-2r}{2}}} G_{\Del, 0}^{(\rs)}(z, \bz)\mid_{\Del_i \to \Del_i +\sum_{j}k_{ij}}.
\ee
This is consistent with the observation made in \cite{GWD1} (see (5.37) there), where the authors considered the holographic dual configuration to a d-dimensional conformal partial wave, namely ``geodesic Witten diagram'', and pointed out a simple relation between the scalar and general tensor exchange cases. 
Essentially the Gegenbauer polynomial arising from the restriction of spin $J$ propagator along the geodesics can only depend on the external coordinates $\{x_{ij}^2\}$ and geodesic parameters.
After the integration along the geodesic parameters, the functional form of individual terms in the integrand does not change, only the external scaling dimensions are shifted by integer units which are determined by the multinomial expansion. 
%%%%%%%%%%%%%%%%%%%%%%%
\paragraph{}
We should stress here that an explicit expression for general $d$-dimensional conformal block with integer spin $J$ was already available in terms of a double infinite series expansion of Gegenbauer polynomials \cite{DO-2003} (See also \cite{Schomerus1} for the connection with Harish-Chandra functions of $BC_2$ Calogero-Sutherland system), 
where the expansion coefficients can be determined recursively through the quadratic Casimir equation. 
Our result \eqref{Gblock-s} provides an alternative explicit expression in terms of a {\it finite} summation over Appell's hypergeometric functions $\bF_4$ which will facilitate 
the crossing kernel computation through its Mellin-Barnes representation \eqref{Def:tF4} later. 
%\paragraph{}
It would be highly non-trivial to demonstrate the equivalence of the two expressions, to this end we can first consider the following non-trivial identity between ${\bF}_4$ functions and the so-called zonal generalization of hypergeometric function \cite{Koornwinder-1}:
\bea\label{3d-Id1}
&&
{}_2{{F}}_1 
 \left[
\begin{matrix}
~a,~b\\
~c
\end{matrix};
\left[\begin{matrix}
~z,~0\\
~0,~\bz
\end{matrix}\right]
\right] 
=\nn\\
&&\small
{\frac{\Ga(c)}{\Ga(\frac{3}{2}-c)\Ga(a)\Ga(b)\Ga(c-a)\Ga(c-b)}\left(\tilde{\bf{F}}_4 
 \left[
\begin{matrix}
~a,~b\\
~c-\frac{1}{2},~1+a+b-c
\end{matrix}; \ru, \rv
\right]
+\rv^{c-a-b} 
\tilde{\bf{F}}_4 
 \left[
\begin{matrix}
~c-a,~c-b\\
~c-\frac{1}{2},~1-a-b+c
\end{matrix}; \ru, \rv
\right]
\right)},\nn\\
\eea
where $(\ru, \rv)$ and $(z,\bz)$ are related as in \eqref{Def:CR1} and the two variable zonal generalization of ${}_2 F_1$ is explicitly given in terms of the Legendre polynomial $P_n(x)$:
\be\label{3d-Id2}
{}_2{{F}}_1 
 \left[
\begin{matrix}
~a,~b\\
~c
\end{matrix};
\left[\begin{matrix}
~z,~0\\
~0,~\bz
\end{matrix}\right]
\right] =\sum_{m=0}^\infty \sum_{l=0}^m \frac{(a)_m(a-\frac{1}{2})_l (b)_m (b-\frac{1}{2})_l (\frac{3}{2})_{m-l}}{l!(c)_m(c-\frac{1}{2})_l (\frac{3}{2})_m  (\frac{1}{2})_{m-l}} \ru^{\frac{m+l}{2}}P_{m-l}\left(\sigma\right).
\ee
This implies that in our case if we set $d=3$, $a =\kappa_{\sigma_s}^{2 (\rs)}+r$, $b=\kappa_{\sigma_s}^{3 (\rs)}+r$ 
and $c =\frac{3}{2}+i\sigma_s\nu$, we can express the three dimensional scalar conformal partial waves in terms of an infinite summation over Legendre polynomials.
More explicitly we have:
\bea\label{Psi-s-full-3d}
&&G^{(\rs)}_{\Del, J}(z, \bz) 
=\frac{1}{\bc^{(\rs)}_{\Del, J}}
\widetilde{\sum_{r, k}}\prod_{i=1}^4 \frac{ \Ga\left(\kappa^{i (\rs)}_+(\tau)+r\right) }{\Ga\left(\gamma^{(\rs)}_{i}(\tau)-J+r+\sum_{j}k_{ji}\right)}\ru^{\frac{\Del-J}{2}+r} \rv^{k_{23}} \sum_{m=0}^\infty \sum_{l=0}^m
(-1)^l 
\nn\\
&&\times 
 \frac{(\frac{3}{2}+m)_{-l} \Ga(h-\Del-l)}{ l!  (\frac{1}{2})_{m-l}} \frac{(\kappa_+^{2 (\rs)}(\tau)+r, \kappa_+^{3 (\rs)}(\tau)+r)_m (\kappa_+^{2 (\rs)}(\tau)+r-\frac{1}{2}, \kappa_+^{3 (\rs)}(\tau)+r-\frac{1}{2})_l }{\Ga(\Del+m)} \ru^{\frac{m+l}{2}}C_{m-l}^{(\frac{1}{2})}\left(\sigma\right),\nn\\
\eea
and notice that we have used the fact $C_{n}^{(\frac{1}{2})}(\sigma) = P_n(\sigma)$ for $h =\frac{3}{2}$. 
It would be interesting to perform the explicit summation over $\{r, k_{ij}\}$ to demonstrate their equivalence.
Finally, it is useful to note that at the so-called crossing symmetric point $z =\bz =\frac{1}{2}$ or $\ru = \rv =\frac{1}{4}$ which is popular in conformal bootstrap computations, we can reduce to a generalized hypergeometric function:
\be\label{Reduction}
 {\bf{F}}_4   \left[
\begin{matrix}
~a_1,~a_2\\
~c_1,~c_2
\end{matrix}; \ru, \ru
\right] = {}_4 F_3  \left[
\begin{matrix}
~a_1,~a_2,~\frac{c_1+c_2-1}{2},~\frac{c_1+c_2}{2}\\
~c_1,~c_2,~c_1+c_2-1
\end{matrix}; 4\ru
\right],
\ee
this is much easier to be implemented numerically than the ${\bf{F}}_4$ function.
%%%%%%%%%%%%%%%%%%%%%%%%%%%
\paragraph{}
Starting with the $s$-channel conformal partial wave,  we can obtain other equivalent  $t$- and $u$- exchange channels by performing the crossing transformations:
\be\label{CrossingTrans1}
\rs\leftrightarrow \rt~:~(x_2, \Del_2) \leftrightarrow (x_4, \Del_4), \quad \rs\leftrightarrow \ru~:~(x_2, \Del_2) \leftrightarrow (x_3, \Del_3),
\ee
or in terms of the conformally invariant cross ratios \eqref{Def:CR1},
the crossing transformations act on them as:
\be\label{CrossingTrans2}
\rs\leftrightarrow \rt~:~(\ru, \rv) \to (\rv, \ru);~(z,\bz) \to (1-z, 1-\bz), \quad \rs\leftrightarrow \ru~:~(\ru, \rv) \to ({1}/{\ru}, {\rv}/{\ru});~(z,\bz) \to ({1}/{z}, {1}/{\bz}).
\ee
By performing the crossing transformations \eqref{Def:CR1}, we can write down the corresponding conformal partial waves in the $t$- and $u$-channels: 
\bea
\Psi^{(\rt)}_{\nu, J}(x_i) 
&=&\cT^{(\rt)}_{\Del_i}(x_i)
\sum_{\sigma_t =\pm}
\widetilde{\sum_{r, k}}\prod_{i=1}^4\frac{1}{\Ga\left(\gamma^{(\rt)}_{i}-J+r+\sum_{j}k_{ji}\right)} \frac{ \rv^{\tau_{\sigma_t }+r} }{ \ru^{\frac{\ra^{(\rt)}+\rb^{(\rt)}-k_{12}-k_{34}}{2}}}
\nn\\
&\times&
{\left[ \ru^{\frac{\varpi_{12}^{(\rt)}}{2}} \tilde{\bf{F}}_4 
 \left[
\begin{matrix}
~\kappa_{\sigma_t}^{1(\rt)}+r,~\kappa_{\sigma_t}^{2(\rt)}+r\\
~1+i\sigma_t\nu,~1+\varpi_{12}^{(\rt)}
\end{matrix}; \rv, \ru
\right]
+\ru^{\frac{\varpi^{(\rt)}_{34}}{2}} \tilde{\bf{F}}_4 
 \left[
\begin{matrix}
~\kappa_{\sigma_t}^{3(\rt)}+r,~\kappa_{\sigma_t}^{4(\rt)}+r\\
~1+i\sigma_t\nu,~1+\varpi_{34}^{(\rt)}
\end{matrix}; \rv, \ru
\right]
\right]},\nn\\
\label{Psi-t-full}
\eea
%%%%%%%%%%%%%%%%%%%%%%%%%%%%
\bea
\Psi^{(\ru)}_{\nu, J}(x_i) &=&\cT^{(\ru)}_{\Del_i}(x_i)
\sum_{\sigma_u =\pm}
\widetilde{\sum_{r, k}}
\prod_{i=1}^4\frac{1}{\Ga\left(\gamma^{(\ru)}_{i}-J+r+\sum_{j}k_{ji}\right)} 
 \frac{ (\frac{1}{\ru})^{\tau_{\sigma_u }+r} }{ (\frac{\rv}{\ru})^{\frac{\ra^{(\ru)}+\rb^{(\ru)}-k_{14}-k_{23}}{2}}}
\nn\\
&\times&
{\left[ \left(\frac{\rv}{\ru}\right)^{\frac{\varpi_{14}^{(\ru)}}{2}} \tilde{\bf{F}}_4 
 \left[
\begin{matrix}
~\kappa_{\sigma_u}^{1(\ru)}+r,~\kappa_{\sigma_u}^{4(\ru)}+r\\
~1+i\sigma_u\nu,~1+\varpi^{(\ru)}_{14}
\end{matrix}; \frac{1}{\ru}, \frac{\rv}{\ru}
\right]
+\left(\frac{\rv}{\ru}\right)^{\frac{\varpi_{23}^{(\ru)}}{2}} \tilde{\bf{F}}_4 
 \left[
\begin{matrix}
~\kappa_{\sigma_u}^{2(\ru)}+r,~\kappa_{\sigma_u}^{3(\ru)}+r\\
~1+i\sigma_u\nu,~1+\varpi_{23}^{(\ru)}
\end{matrix}; \frac{1}{\ru}, \frac{\rv}{\ru}
\right]
\right]}.\nn\\
\label{Psi-u-full}
\eea
Here we have used the crossing transformations \eqref{CrossingTrans2} to change the cross ratios, 
while various remaining parameters $(\kappa_{\sigma_t}^{i(\rt)}, \gamma_{i}^{(\rt)}; \kappa_{\sigma_u}^{i(\ru)}, \gamma_{i}^{(\ru)})$
and pre-factors $(\cT^{(\rt)}_{\Del_i}(x_i), \cT^{(\ru)}_{\Del_i}(x_i))$ can be obtained from \eqref{gamma0i-s} and \eqref{Def:kappa-s} using the crossing transformations \eqref{CrossingTrans1}. It is interesting to note that the $s$ and $t$-channel conformal partial waves are both convergent within the square given by $0< \ru, \rv <1$ 
or equivalently $0< z, \bz <1$, this becomes important when we compute the crossing kernel, as we need to use the expressions for the integrand which are convergent in a given integration region.
%%%%%%%%%%%%%%%%
\section{Generalization to Continuous Spins}
\paragraph{}
For our computation of the crossing kernel using the inversion formula of \cite{CH-2017} in the next section, here we would also like to consider the scalar conformal block in arbitrary dimensions with continuous spin $J$ by generalizing our earlier construction.
Recall that the Mellin representation for conformal  partial wave with integer $J$ \eqref{Psi-s-Mellin1} can be derived from pairing two copies of three point functions related through a shadow transformation and the contraction of their symmetric traceless tensor structures yields a Gegenbauer polynomial $C_{J}^{(\veps)}(\eta)$ in the process, see for example \cite{DO-2011}, \cite{CKK2}.
Now for continuous spin $J$, we need to generalize the shadow to ``full-shadow'' transformation \cite{LightRay}, this involves contracting two copies of three point functions whose spins are also related via the spin shadow transformation \cite{Work1}.
The net result here is to replace the Gegenbauer polynomial arising from the contraction of two symmetric traceless tensors with identical integer spin $J$:
\be\label{IntJ-contraction}
\hat{X}^{\{\mu_1\dots}\hat{X}^{\mu_J\}} \hat{Y}_{\{\mu_1\dots}\hat{Y}_{\mu_J\}} = \frac{1}{2^J c_J} {}_2F_1
\left[\begin{matrix}
~-J,~J+2\veps\\
\veps+\frac{1}{2}
\end{matrix}
; \frac{1-\eta}{2}
\right] = \frac{1}{2^J c_J}\hat{C}_{J}^{(\veps)}(\eta),~~\eta = {\hat{X}\cdot \hat{Y}}
\ee
where $\hat{X}, \hat{Y}$ are unit vectors and $\hat{C}_{J}^{(\veps)}(\eta) =\frac{J!}{(2\veps)_J} C_{J}^{(\veps)}(\eta)$,
by its continuous spin generalization through integration over the null polarization vector $z^\nu$ \cite{LightRay}:
\bea\label{Cont-Gegenbauer}
&&\int D^{2\veps} z (-2z\cdot \hat{X})^{J} (-2z\cdot \hat{Y})^{\bJ} 
= \pi^{\veps}\frac{\Ga(\veps)}{\Ga(2\veps)} {}_2F_1
\left[\begin{matrix}
~-J,~-\bJ\\
\veps+\frac{1}{2}
\end{matrix}
; \frac{1-\eta}{2}
\right],\nn\\
&&=\pi^{\veps}\left\{\frac{\Ga(J+\veps)}{\Ga(J+2\veps)} (2\eta)^J {}_2 F_1 \left[\begin{matrix}
~-\frac{J}{2},~\frac{1-J}{2}\\
2-h-J
\end{matrix}
; \frac{1}{\eta^2}
\right]+ \frac{\Ga(\bJ+\veps) }{\Ga(\bJ+2\veps)}(2\eta)^{\bJ} {}_2 F_1 \left[\begin{matrix}
~-\frac{\bJ}{2},~\frac{1-\bJ}{2}\\
2-h-\bJ
\end{matrix}
; \frac{1}{\eta^2}
\right]\right\}.\nn\\
\eea
Here we have introduced the shadow spin $\bJ = -2\veps-J$ and the scale-invariant integration measure over polarization vector $z^\mu$:
\be\label{Def:Spin-Measure}
D^{2\veps}z \equiv \frac{d^d z \theta(z^0)\delta(z^2)}{{\rm vol} {\mathbb{R}}_+}.
\ee
where the division by the volume of  the group of positive rescaling ${{\rm vol} {\mathbb{R}}_+}$ ensures the finite result.
In the second equality of \eqref{Cont-Gegenbauer}, we have separated the dependence on continuous spin $J$ and its shadow $\bJ$ using hypergeometric identity.
It is clear that unless in even dimensions, i.e. $\veps =0, 1, 2, \dots $ etc.,  
the $\Gamma$-functions in the pre-factors ensure that when $J \in {\mathbb{Z}}_{\ge 0}$ or equivalently $\bJ+2\veps \in {\mathbb{Z}}_{\le 0}$, the $\bJ$-dependent term vanishes identically and vice versa.
\paragraph{}
The rest of the computation proceeds almost identically as in the integer spin case, 
the Mellin representation for the conformal partial wave with continuous spin is now given by two distinct spin-dependent parts:
\bea\label{Psi-s-ConJ}
\tilde{\Psi}_{\nu, \ell}^{(\rs)}(x_i) &=& 
\cT_{\Del_i}^{(\rs)}(x_i) \int^{i\infty}_{-i\infty} \frac{ds}{(4\pi i)}  \int^{i\infty}_{-i\infty} \frac{dt}{(4\pi i)} \ru^{\frac{s}{2}} \rv^{\frac{t}{2}}  \rho^{(\rs)}_{\Del_i} (s, t)\left( \frac{\Ga(+i\ell)}{\Ga(\veps+i\ell)}\bM^{(\rs)}_{\nu, +\ell}(s, t) + \frac{\Ga(-i\ell)}{\Ga(\veps-i\ell)}\bM^{(\rs)}_{\nu, -\ell}(s, t)\right)\nn\\
&=&\cT_{\Del_i}^{(\rs)}(x_i) \sum_{\sigma_s = \pm }\sum_{\eta_s =\pm}\hat{\bc}_{h+i\sigma_s \nu, -\veps+i\eta_s\ell}  \frac{\Ga(i\eta_s \ell)}{\Ga(\veps+i\eta_s \ell)} \fg_{h+i\sigma_s\nu, -\veps+i\eta_s\ell }^{(\rs)}(z, \bz) .
\eea
Here we have also parameterized continuous spin $J = -\veps+i\ell$ and shadow spin $\bJ = -\veps-i\ell$ through spectral parameter $\ell$, putting it on equal footing as the scaling dimension.
In the last line of \eqref{Psi-s-ConJ}, we have made both shadow and spin-shadow transformations manifest after $(s, t)$-integration by expressing them in terms of the building blocks 
$ \fg_{h+i\sigma_s\nu, -\veps+i\eta_s\ell }^{(\rs)}(z, \bz)$ whose explicit form will be given momentarily. Here the normalization constants:
\be\label{Cont-coeff}
\hat{\bc}_{h+i\sigma_s \nu, -\veps+i\eta_s \ell} = {2^{-\veps+i\eta_s \ell}}\pi^{\veps} \frac{\Ga(-i \sigma_s\nu)\Ga(-i \sigma_s \nu+i\eta_s \ell)}{\Ga(2\ow_{\sigma_s \eta_s})\Ga(h-i\sigma_s \nu-1)}\frac{ \Ga\left(\ow_{\sigma_s\eta_s}\pm \ra^{(\rs)}\right)  
\Ga\left(\ow_{\sigma_s\eta_s}\pm\rb^{(\rs)}\right)}{\prod_{i=1}^4 \Ga\left(\ga_{i  \eta_s}^{(\rs)}\right)}.
\ee
are fixed by comparing \eqref{Cont-Gegenbauer} with \eqref{IntJ-contraction} and \eqref{Gblock-s}, 
such that $J$ becomes an non-negative integer, we recover the conformal block given in \eqref{Gblock-s} from $\fg_{\Del, J}^{(\rs)}(z, \bz)$.
We can also define the following parameters:
\be
\tau_{\sigma_s \eta_s} = \frac{h+i\sigma_s \nu - (-\veps+ i\eta_s \ell)}{2},\quad 
\ow_{\sigma_s \eta_s} = \frac{h+i\sigma_s \nu + (-\veps +i\eta_s \ell)}{2}, \quad \sigma_s, \eta_s = \pm,
\ee
\be\label{Def:gammai+}
\ga_{1\eta_s}^{(\rs)} = \ow_{+ \eta_s}-\ra^{(\rs)},~ \ga_{2 \eta_s}^{(\rs)} =\ow_{+ \eta_s} +\ra^{(\rs)},~
\ga_{3  \eta_s}^{(\rs)} = \ow_{- \eta_s}+\rb^{(\rs)},~ \ga_{4 \eta_s}^{(\rs)} =\ow_{- \eta_s} -\rb^{(\rs)}.
\ee
\bea
\label{Def:kappa-s-cont}
&&\kappa_{\sigma_s\eta_s}^{1 (\rs)} = \tau_{\sigma_s\eta_s}-\ra^{(\rs)}+k_{13}+k_{14},~ \kappa_{\sigma_s\eta_s}^{2 (\rs)} =  \tau_{\sigma_s\eta_s} +\ra^{(\rs)}+k_{23}+k_{24},\nn\\
&&\kappa_{\sigma_s\eta_s}^{3 (\rs)}=   \tau_{\sigma_s\eta_s}+\rb^{(\rs)}+k_{13}+k_{23} ,~ \kappa_{\sigma_s\eta_s}^{4 (\rs)} = \tau_{\sigma_s\eta_s} -\rb^{(\rs)}+k_{14}+k_{24}.
\eea
and the precise definition of non-integral summation indices $\{k_{ij}\}$ is given around \eqref{Sum-k}.
We can now write down ${\bf{M}}^{(\rs)}_{\nu, \eta_s \ell}(s, t)$, each contains two pieces related by shadow transformation after the Mellin variable $(s, t)$ integrations:
\bea
&&{\bf{M}}^{(\rs)}_{\nu, \eta_s \ell}(s, t) = \frac{\pi^{\veps}}{\prod_{i=1}^4 \Ga\left(\ga_{i\eta_s}^{(\rs)}\right)} \frac{\Ga\left(\tau_{\pm \eta_s}-\frac{s}{2}\right)}{\Ga(\delta_{12}^{(\rs)})\Ga(\del_{34}^{(\rs)})}
\tQ_{\nu,\eta_s \ell}^{(\rs)}(s,t),\label{Cont-Mellin}\\
&&\tQ_{\nu,\eta_s \ell}^{(\rs)}(s,t)=
\widetilde{\sum_{m, k}}
\left(\tau_{\pm \eta_s}-\frac{s}{2}\right)_m
\prod_{(ij)}\frac{\Ga\left(\delta_{ij}^{(\rs)}+k_{ij}\right)}{\Ga\left(\delta_{ij}^{(\rs)}\right)} \prod_{i=1}^4\frac{ \Ga\left(\ga_{i \eta}^{(\rs)}\right) }{\Ga\left(\ga_{i\eta_s}^{(\rs)}-(-\veps+i\eta_s \ell)+m+\sum_{j}k_{ij}\right) }.
\label{Mack-Q}\nn\\
\eea
We can regard  $\tQ_{\nu,\eta\ell}^{(\rs)}(s,t)$ as a continuous spin generalization of a Mack polynomial, here $\{\del_{ij}^{(\rs)}\}$ containing Mellin variables are parameterized as before \eqref{deltaij}.
The abridged infinite summation is given by
:
\be\label{Def:Summation-Cont}
\widetilde{\sum_{m, k}} \dots \equiv \sum_{m=0}^{\infty} 2^{2m}\frac{(\frac{\veps-i\eta_s\ell}{2})_m(\frac{h-i\eta_s\ell}{2})_m}{m!(1-i\eta_s\ell)_m} 
 \widetilde{\sum_{\sum k_{ij}=-\veps+i\eta_s\ell-2m}} (-1)^{k_{13}+k_{24}}\frac{ \Ga(-\veps+i\eta_s\ell-2m +1) }{\prod_{(ij)}\Ga(k_{ij}+1) } \dots,
\ee
where the first infinite summation over $m$ comes from the expansion of a ${}_2F_1$ function, while the remaining summation comes from analytically continuing the multinomial 
theorem to continuous power \cite{Hilliker}, such that it is also an infinite summation:
\be\label{Sum-k}
\widetilde{\sum_{\sum k_{ij} \dots = -\veps+i\eta_s\ell-2m}} \equiv \sum_{p=0}^\infty \sum_{k_{14}+k_{24}+k_{13} = p} \dots,~~k_{24} = -\veps+i\eta_s \ell-2m-p.
\ee
In other words, we generalize the earlier four-fold partition of integer spin by picking one of the four $k_{(ij)}$, say $k_{24}$ to be a continuous parameter.
Altogether, as in the integer spin case, we now have five (non-independent) summations, even though the range now expands to infinities to account for continuous spin.
\paragraph{}
To obtain the conformal block in general $d$-dimension with continuous spin $G_{\Del, J}^{(\rs)}(z, \bz)$, we can explicitly integrate over Mellin variables $(s, t)$ as before to obtain the building block $\fg_{\Del, J}^{(\rs)}(z, \bz)$ and $\fg_{\Del, \bJ}^{(\rs)}(z, \bz)$ in terms of  Appell's hypergeometric function $F_4$, as the pole structures remain the same here. We can then consider the following linear combination\footnote{Notice that we could have chosen slightly more symmetric combination such as:
\be
G_{\Del, J}^{(\rs)}(z, \bz) \sim \frac{\Ga(J+\veps)}{\Ga(J+2\veps)} \fg_{\Del, J}^{(\rs)}(z, \bz) + \frac{\Ga(\bJ+\veps)}{\Ga(\bJ+2\veps)} \fg_{\Del, \bJ}^{(\rs)}(z, \bz),
\ee
however the normalization of the conformal block will need to change accordingly.
 }:
\bea\label{Gblock-s-cont}
&&G^{(\rs)}_{\Del, J}(z, \bz)
=  \fg_{\Del, J}^{(\rs)}(z, \bz) + \frac{\Ga(J+2\veps)\Ga(\bJ+\veps)}{\Ga(\bJ+2\veps)\Ga(J+\veps)} \fg_{\Del, \bJ}^{(\rs)}(z, \bz),
\eea
where the building blocks are given explicitly by
\bea\label{Def:fgblock}
&&\fg_{\Del, J}^{(\rs)}(z, \bz) = \frac{\pi^\veps}{\hat{\bc}_{\Del, J}}\widetilde{\sum_{m, k}} \prod_{i=1}^4\frac{ 1}{\Ga\left(\ga_{i+}^{(\rs)}(\Del, J)-J+m+\sum_{j}k_{ij}\right) } \frac{ \ru^{\frac{\Del-J}{2}+m} }{ \rv^{\frac{\ra^{(\rs)}+\rb^{(\rs)}-k_{14}-k_{23}}{2}}}
\nn\\
&&\times
{\left[ \rv^{\frac{\varpi_{14}^{(\rs)}}{2}} \tilde{\bf{F}}_4 
 \left[
\begin{matrix}
~\kappa_{++}^{1(\rs)}(\Del, J)+m,~\kappa_{++}^{4(\rs)}(\Del, J)+m\\
~1-h+\Del,~1+\varpi_{14}^{(\rs)}
\end{matrix}; \ru, \rv
\right]
+\rv^{\frac{\varpi_{23}^{(\rs)}}{2}} \tilde{\bf{F}}_4 
 \left[
\begin{matrix}
~\kappa_{++}^{2(\rs)}(\Del, J)+m,~\kappa_{++}^{3(\rs)}(\Del, J)+m\\
~1-h+\Del,~1+\varpi_{23}^{(\rs)}
\end{matrix}; \ru, \rv
\right]
\right]},\nn\\
\eea
while $\varpi_{14}^{(\rs)}$ and $\varpi_{23}^{(\rs)}$ remain the same in \eqref{vw14} and \eqref{vw23}.
Here $\ga_{i+}^{(\rs)}(\Del, J)$ and $\kappa_{++}^{i(\rs)}(\Del, J)$ are given by setting $h+i\nu = \Del$ and $-\veps+i\ell =J$ in $\ga_{i+}^{(\rs)}$ and $\kappa_{++}^{i(\rs)}$ defined in  \eqref{Def:gammai+} and \eqref{Def:kappa-s-cont}, while $\fg_{\Del, \bJ}^{(\rs)}(z, \bz)$ can be obtained by obvious $J \to \bJ$ replacements in the parameters, such as setting $h+i\nu = \Del$ and $-\veps-i\ell =\bJ$ in $\ga_{i-}^{(\rs)}$ and $\kappa_{+-}^{i(\rs)}$ to obtain $\ga_{i-}^{(\rs)}(\Del, \bJ)$ and $\kappa_{+-}^{i(\rs)}(\Del, \bJ)$ respectively.
One should keep in mind that one of $\{k_{(ij)}\}$ is now a non-integer, we should perform infinite summation as in \eqref{Sum-k} instead.
We have numerically checked that our expression matches with the known results for the conformal blocks with continuous spin in $d=2, 4$ dimensions, i.e. directly taking $J$ to be continuous in the argument of hypergeometric functions as was done in \cite{Liu-2018}.
\paragraph{}
We should note previously the explicit form of scalar conformal block with continuous spin $J$ was also constructed in \cite{Schomerus1}, where the authors 
constructed the following linear combination:
\be\label{Cont-block-IS}
G_{\Del, J}^{(\rs) {\rm IS}}(z,\bz) =\frac{(\veps)_J}{(2\veps)_J}\hat{\Phi}_{\Del, J}^{(\rs)}(z, \bz) + \frac{(\veps)_{\bJ}}{(2\veps)_{\bJ}} \hat{\Phi}_{\Del, \bJ}^{(\rs)}(z, \bz),~~
\hat{\Phi}_{\Del, J}^{(\rs)}(z, \bz) = \frac{2^{4\ra^{(\rs)}+2\Del}}{(z\bz)^{\ra^{(\rs)}}} \tilde{\Phi}_{\Del, J}^{(\rs)}(z, \bz).
\ee
Here $\tilde{\Phi}_{\Del, J}^{(\rs)}(z, \bz)$ is the so-called twisted Harish-Chandra function which is related to the eigenfunction of the $BC_2$ Calogero-Sutherland Hamiltonian via a similarity transformation. The explicit form of $\tilde{\Phi}_{\Del, J}^{(\rs)}(z, \bz)$ depends on $(\Del, J, \ra^{(\rs)}, \rb^{(\rs)}, h)$ and can be written in terms of a double infinite series involving ${}_2F_1$. While in \cite{CH-2017}, it was proposed that the scalar conformal block with continuous $J$ exchange can be decomposed as
\be\label{Cont-block-CH}
G_{\Del, J}^{(\rs) {\rm CH}}(z,\bz) = g^{\rm pure}_{\Del, J}(z, \bz) + \frac{\Ga(J+2\veps)\Ga(\bJ+\veps)}{\Ga(\bJ+2\veps)\Ga(J+\veps)} g^{\rm pure}_{\Del, \bJ}(z, \bz),
\ee
where the function $g_{\Del, J}^{\rm pure}(z, \bz)$ in \cite{CH-2017}, which is also a solution of the quadratic Casimir equation \eqref{Casimir2-eqn}, 
but with different asymptotic behavior from $G_{\Del, J}^{(\rs)}(z, \bz)$, i.e.:
\bea
G_{\Del, J}^{(\rs) {\rm IS/CH}} (z, \bz) &\to& ({\rm{Const.}}) (z\bz)^{\frac{\Del}{2}}{}_2 F_1 \left[\begin{matrix}
~-J,~-\bJ\\
\veps+\frac{1}{2}
\end{matrix}
; \frac{1-\sigma}{2}
\right]+\cO\left((z\bz)^{\frac{\Del}{2}+1}\right),\label{Gcont-asymp}
\\
g^{\rm pure}_{\Del, J}(z, \bz) &\to& (z\bz)^{\frac{\Del}{2}} (2\sigma)^J {}_2 F_1 \left[\begin{matrix}
~-\frac{J}{2},~\frac{1-J}{2}\\
2-h-J
\end{matrix}
; \frac{1}{\sigma^2}
\right]+\cO\left((z\bz)^{\frac{\Del}{2}+1}\right),\quad z\bz\to 0,~\sigma = \frac{z+\bz}{2\sqrt{z\bz}}~~{\rm fixed}. \label{gpure-asymp}\nn\\
\eea
The two distinct asymptotic behaviors \eqref{Gcont-asymp} and \eqref{gpure-asymp} can be reconciled using the hypergeometric function identity given in \eqref{Cont-Gegenbauer}, 
this also implies that \eqref{Cont-block-IS} and \eqref{Cont-block-CH} can be directly matched if $\hat{\Phi}_{\Del, J}^{(\rs)}(z, \bz) $ and $g^{\rm pure}_{\Del, J}(z, \bz)$ are identified up to an overall constant factor, 
as it was indeed done in \cite{Schomerus1}.
We have also numerically checked asymptotic behavior of our expression for continuous spin conformal block \eqref{Gblock-s-cont}, which also matches with \eqref{Gcont-asymp} up to an overall constant for continuous $(\Del, J)$. This implies that our expression $\fg_{\Del, \bJ}^{(\rs)}(z, \bz)$ \eqref{Def:fgblock} provides an alternative expression for the twisted Harish Chandra function $\hat{\Phi}_{\Del, \bJ}^{(\rs)}(z, \bz)$ hence $g^{\rm pure}_{\Del, J}(z, \bz)$ as derived from the corresponding Mellin amplitude \eqref{Cont-Mellin} and \eqref{Mack-Q}.
It would be interesting but non-trivial to analytically demonstrate the full functional equivalence of  \eqref{Def:fgblock} and $\hat{\Phi}_{\Del, \bJ}^{(\rs)}(z, \bz)$ using appropriate hypergeometric function identities, while it seems obvious to us they should match, as they both solve the quadratic Casimir equation and their asymptotic behaviors numerically match.
%%%%%%%%%%%%%%%%%%%%%%%%%%%%%
%%%%%%%%%%%%%%%%%%%%%%%%%%%%%
\section{Lorentzian Crossing Kernel and Kamp\'{e} de F\'{e}riet Functions}\label{Sec:Crossing}
\paragraph{}
In the Euclidean space ${\mathbb{R}}^d$, the conformal partial wave $\Psi_{\nu, J}^{(\rs)}(x_i)$ \eqref{Psi-s-full} for integer spin $J$ is known to satisfy the orthogonality condition \cite{CH-2017, SSW-2017} with respect to the following inner product:
\be\label{Ortho-cond1}
\left( \Psi^{(\rs)}_{\nu, J}(x_i), {\Psi^{(\rs)}_{\nu', J'}}(x_i) \right) = \int\frac{\prod_{i=1}^4 d^d x_i}{{\rm Vol}(SO(1,d+1))} \Psi^{(\rs)}_{\nu, J}(x_i) \overline{\Psi}^{(\rs)}_{\nu', J'}(x_i)=  2\pi{\cN_{\nu, J}}\left(\delta(\nu-\nu')+\cK_{\nu}\, \delta(\nu+\nu')\right)\delta_{J, J'},
\ee
where
$\cN_{\nu, J}$ is the overall normalization factor\footnote{Here we take the overall normalization factor to be:
\be\label{Norm-Factor}
\cN_{\nu, J} =\lambda_0^2 \frac{\pi^{d+1}}{2^{2(J+\veps)}}\frac{{\rm vol}(S^{d-2})}{{\rm vol}(SO(d-1))}\frac{(2J+2\veps)\Gamma(J+1)\Gamma(J+2\veps)}{\Gamma(J+h)^2}\frac{\Ga(\pm i\nu)}{\Ga(h\pm i\nu-1)\Ga(h\pm i\nu-J-1)}.
\ee
where $\lambda_0 = \frac{\Psi_{\nu, J}^{(\rs)}(x_i)}{\Psi_{\Del, J}^{\Del_i}(x_i)}$ is the ratio of the conformal partial waves between \eqref{Def:Psinu1} and the one  used in \cite{SSW-2017}, which takes care of difference in normalization conventions.}.
The factor $\cK_{\nu}$ arises from the property of CPW: $\Psi_{-\nu,J}=\frac{1}{\cK_{\nu}} \Psi_{\nu,J}$ and we will obtain the expression of this coefficient from the bulk calculation as in \eqref{coef: K}.
Here ``$\overline{{\Psi}}$'' denotes performing the shadow transformation on all scaling dimensions involved, i.e.  $\{\Del_i, h+i\sigma_s \nu'\} \to \{d-\Del_i, h-i\sigma_s\nu'\}$,
and $\delta_{J, J'}$ arises from the orthogonality of the Gegenbauer polynomial $C_{J}^{(\veps)}(x)$. The integration over $\{x_i\}$ is necessarily divided by the volume of Euclidean conformal symmetry group ${\rm Vol}(SO(1, d+1))$.
We will provide the an alternative derivation for this orthogonality condition by considering the harmonic function in Euclidean AdS space \cite{GWD1} shortly in Section \ref{Sec:AdS}.
\paragraph{} 
As the s-channel conformal partial waves $\Psi^{(\rs)}_{\nu, J}(x_i)$ form an orthogonal basis, it is interesting to ask whether we can expand a conformal partial wave in other exchange channels
in terms of them. To obtain the relevant expansion coefficients for t-channel conformal partial wave $ \Psi^{(\rt)}_{\nu, J}(x_i)$, we naturally consider the following integral:
\be\label{st-integral}
\left( \Psi^{(\rt)}_{\nu, J}(x_i),{\Psi^{(\rs)}_{\nu', J'}}(x_i) \right) = 
\int\frac{\prod_{i=1}^4 d^d x_i}{{\rm Vol}(SO(1,d+1))} \Psi^{(\rt)}_{\nu, J}(x_i) \overline{\Psi}^{(\rs)}_{\nu', J'}(x_i) ,
\ee
this is an example of what sometimes referred as (Euclidean) ``crossing kernel'' \cite{CrossingKernel-1} or ``$6j$ symbol'' \cite{Liu-2018} (See also \cite{Karateev-6J} for earlier work),
we will give its more explicit definition, i. e. in terms of appropriate spectral parameters momentarily.
\paragraph{}
To perform the integral in the \eqref{st-integral}, we can rewrite the integration measure in terms of the cross ratios $(z, \bz)$.
However in Euclidean signature, the natural integration range for $(z, \bz)$ spans the entire complex plane ${\mathbb{C}}$, this requires us to consider the analytic continuation of $\bF_4$ or infinite summation 
of ${}_2F_1$s around their branch points, these occur when pairs of external points coincide, i.e. $z=\bz=0,1, \infty$. 
This seemingly complicated problem was greatly simplified, as deduced first in \cite{CH-2017} and later in \cite{SSW-2017}, if we gauge fix some of the coordinates and wick-rotate the remaining ones into Lorentzian signature.
In Lorentzian spacetime signature, $(z, \bz)$ are no longer related by complex conjugation but rather independent variables, 
we can now consider the singularities and the associated monodromies for each variable independently. The detailed analysis in \cite{CH-2017} and \cite{SSW-2017} show that the integration range now factorize into different regions and the respective integrands now involve certain ``double discontinuities''. 
The more general results in \cite{CH-2017, SSW-2017} stated that for the inner product involving an arbitrary four point scalar correlation function, we can explicitly evaluate it through the following integrals:
\bea\label{Def:Lorentzian-Inner}
&&\left( \left\langle \prod_{i=1}^4 \cO_{i} (x_i) \right\rangle, \Psi^{(\rs)}_{i(h-\Del), J}(x_i)\right)
=\hat{\alpha}_{\Delta, J}^{(\rs)} \Big[(-1)^{J} \int^1_0 \int^1_0 \frac{dz d\bz}{(z\bz)^d}|z-\bz|^{2\veps}\,G_{\tilde{\Del}, \tilde{J}}^{(\rs)}(z, \bz)\mid_{ \bar{\Del}_i} \frac{\langle[\cO_3, \cO_2][\cO_1, \cO_4]\rangle}{\cT^{(\rs)}_{\Del_i}(x_i)}
\nn\\
&&
+\int^0_{-\infty} \int^0_{-\infty} \frac{dz d\bz}{(z\bz)^d}|z-\bz|^{2\veps}\,\hat{G}_{\tilde{\Del}, \tilde{J}}^{(\rs)}(z, \bz)\mid_{ \bar{\Del}_i } \frac{\langle[\cO_4, \cO_2][\cO_1, \cO_3]\rangle}{\cT^{(\rs)}_{\Del_i}(x_i)}\Big].
\eea
This is sometimes referred as ``Lorentzian inversion formula'', as it can be used to invert the operator product expansion by extracting the corresponding OPE coefficient (up to overall normalization constant) for a given conformal partial wave. Moreover this formula gives the natural analytic continuation of OPE coefficient in spin $J$ to continuous value.
More precisely, the conformal blocks $G_{\tilde{\Del}', \tilde{J}'}^{(\rs)}(z,\bz)\mid_{\bar{\Del}_i}$ and $\hat{G}_{\tilde{\Del}', \tilde{J}'}^{(\rs)}(z,\bz)\mid_{\bar{\Del}_i}$ are now strictly Lorentzian, 
they are associated with the non-local primary operator $\cO_{\tilde{\Del}', \tJ'}$ generated by the spin-shadow, light ray then spin-shadow transformations considered in \eqref{8fold-sym} (sometimes it is called ``flood light transformation'' \cite{LightRay}), with the quantum numbers:\footnote{The overall constant $\hat{\alpha}_{\Delta', J'}^{(\rs)}$ here is given by:
\be
\hat{\alpha}_{\Delta, J}^{(\rs)} = -\lambda_0\frac{a_{\Del, J}}{2^{d+J}}\frac{1}{{\rm vol}(SO(d-1))}\frac{\Ga(J+2\veps)\Ga(\veps)}{\Ga(J+\veps)\Ga(2\veps)}, \quad 
a_{\Del, J}=\frac{(2\pi)^{d-2}}{2}\frac{\Ga(J+1)\Ga(\Del-h)}{\Ga(J+h)\Ga(\Del-1)}\frac{\Ga(\frac{\Del+J}{2}\pm \ra^{(\rs)}) \Ga(\frac{(d-\Del)+J}{2}\pm \rb^{(\rs)})}{\Ga(J+\Del)\Ga(J+\bar{\Del})}.
\ee
where $\lambda_0 = \frac{\Psi_{\nu, J}^{(\rs)}(x_i)}{\Psi_{\Del, J}^{\Del_i}(x_i)}$ takes care of difference in normalizations.
}:
\be\label{FloodLight-param}
{\rm Flood~light}~:~(\Del, J) \to (\tDel=J+d-1, \tJ=\Del-d+1),~{\rm or}~(h+i\nu, -\veps+i\ell)  \to (h+i\ell, -\veps+i\nu),
\ee
such that the spectral parameters $(\nu, \ell)$ in \eqref{Def:Spec2} are also exchanged.
\paragraph{}
To apply \eqref{Def:Lorentzian-Inner} in our subsequent computations, 
we will use our scalar conformal block for continuous spin given in \eqref{Gblock-s-cont} with the transformed parameters \eqref{FloodLight-param}.
The subscript $\mid_{\bar{\Del}_i}$ denotes the external scaling dimensions $\{\Del_i\}$ has been replaced by their shadows $\{\bar{\Del}_i = d-\Del_i\}$.
Moreover $\hat{G}_{\tDel, \tJ}^{(\rs)}(z, \bz)$ is defined as the conformal block such that for negative cross ratios $|z| \ll |\bz|\ll 1$, it scales as $\sim (-z)^{\frac{\tDel-\tJ}{2}}(-\bz)^{\frac{\tDel+\tJ}{2}}$.
The two double commutators  are defined using the appropriate $i\epsilon$ prescription as in computing a Feynman propagator:
\bea
&&\frac{\langle[\cO_3, \cO_2][\cO_1, \cO_4]\rangle}{\cT^{(\rs)}_{\Del_i}(x_i)}=-2 {\rm dDisc}_{\rt}\left[\cF^{(\rs)}(z,\bz)\right]\nn\\
&&=-2\cos\pi\left(\ra^{(\rs)}+\rb^{(\rs)}\right)\cF^{(\rs)}(z, \bz) +e^{i\pi(\ra^{(\rs)}+\rb^{(\rs)})} \cF^{(\rs), {\rm ccw}} (z, \bz) 
+ e^{-i\pi(\ra^{(\rs)}+\rb^{(\rs)})} \cF^{(\rs), {\rm cw}} (z, \bz)\,,\nn\\
\label{double-comm-t}
\\
&&\frac{\langle[\cO_4, \cO_2][\cO_1, \cO_3]\rangle}{\cT^{(\rs)}_{\Del_i}(x_i)}= -2 {\rm dDisc}_{\ru}\left[\cF^{(\rs)}(z,\bz)\right]\nn\\
&&= -2\cos\pi\left(\ra^{(\rs)}-\rb^{(\rs)}\right)\cF^{(\rs)}(z, \bz) +e^{i\pi(\ra^{(\rs)}+\rb^{(\rs)})} \cF^{(\rs), {\rm cw}} (z, \bz) 
+ e^{-i\pi(\ra^{(\rs)}+\rb^{(\rs)})} \cF^{(\rs), {\rm ccw}} (z, \bz)\,,\nn\\
\label{double-comm-u}
\eea
where we have parameterized the four point correlation function as:
\be\label{4pt-Param-s}
\left\langle \prod_{i=1}^4 \cO_{i}(x_i) \right\rangle = \cT^{(\rs)}_{\Del_i}(x_i) \cF^{(\rs)}(z,\bz).
\ee
The superscripts $``{\rm cw}''$ and $``{\rm ccw}''$ in \eqref{double-comm-t} denote if we take $\bz$ around its branch point $\bz=1$ in the clock-wise and counter-clock-wise direction, while keeping $z$ fixed;
while for \eqref{double-comm-u}, we take $\bz = -\infty$ while keeping $z$ fixed.
\paragraph{}
Now returning to the evaluation of \eqref{st-integral} using \eqref{Def:Lorentzian-Inner},
we further restrict $\langle  \cO_{1} \cO_{2} \cO_{3} \cO_{4} \rangle \equiv \Psi^{(\rt)}_{\nu, J}(x_i)$. 
For two and four dimensions, where closed form expressions for Euclidean scalar conformal blocks with integer spins were previously available in \cite{DO-2003}, \cite{DO-2011}, 
this computation was done in \cite{Liu-2018}, and the authors analytically continued the spin $J$ to continuous values, we also showed numerically earlier that such continuations match with our result \eqref{Gblock-s-cont}.
Using the t-channel conformal block \eqref{Psi-t-full}, the double commutator can be readily computed from \eqref{double-comm-t}:
\bea\label{DoubleDisc}
&&-2 {\rm dDisc}_{\rt}\left[\frac{\Psi^{(\rt)}_{\nu, J}(x)}{\cT^{(\rs)}_{\Del_i}(x_i)}\right]=-4 \sum_{\sigma_t =\pm} \sin\pi\left(\tau_{\sigma_t}-\frac{1}{2}\Del_{14}^+\right)  \sin\pi\left(\tau_{\sigma_t}-\frac{1}{2}\Del_{23}^+\right)\bc_{h+i\sigma_t\nu , J}^{(\rt)} \frac{\cT^{(\rt)}_{\Del_i}(x_i)}{\cT^{(\rs)}_{\Del_i}(x_i)} G_{h+i\sigma_t\nu, J}^{(\rt)}(z,\bz),\nn\\
&&-2 {\rm dDisc}_{\ru}\left[\frac{\Psi^{(\rt)}_{\nu, J}(x)}{ \cT^{(\rs)}_{\Del_i}(x_i)}\right]=0.
\eea
The vanishing of the second double discontinuity can also be readily verified by considering the expansion around $\bz =\infty$ and use the identity \eqref{Psi-s-full-3} for t-channel conformal block.
Collecting all the pieces together, we have the following integral expression for the Lorentzian crossing kernel between $s$- and $t$-channels:
\bea\label{CrossingKernel-st-Lorentzian1}
&&\left( \Psi^{(\rt)}_{\nu, J}(x_i),{\Psi^{(\rs)}_{i(h-\Del'), J'}}(x_i) \right) \nn\\
&&= -2{{\hat{\alpha}_{\Delta', J'}^{(\rs)} (-1)^{J'}}} \int^1_0 \int^1_0 \frac{dz d\bz}{(z\bz)^{2}}\left|\frac{z-\bz}{z\bz}\right|^{2\veps}[(1-z)(1-\bz)]^{\ra^{(\rs)}+\rb^{(\rs)}} G_{\tilde{\Del}',\tilde{J}'}^{(\rs)}(z, \bz){\rm dDisc}_{\rt}\left[\frac{\Psi^{(\rt)}_{\nu, J}(x)}{\cT^{(\rs)}_{\Del_i}(x_i)}\right],\nn\\
\eea
where we used the identity \eqref{Shadow-Id} to revert $\{\bar{\Del}_i\}$ to $\{\Del_i\}$ in the $s$-channel conformal block.
The Lorentzian inversion formula now truncates the integration range to $0< z, \bz <1$, where both $s$ and $t$ channel conformal partial waves given in \eqref{Psi-s-full} and \eqref{Psi-t-full} are convergent,
we can therefore use their Mellin-Barnes representations to perform the integration in \eqref{CrossingKernel-st-Lorentzian1}.
We can now parametrize \eqref{st-integral} in terms of the following summation:
\bea\label{CrossingKernel-st-Lorentzian2}
&&\left( \Psi^{(\rt)}_{\nu, J}(x_i),{\Psi^{(\rs)}_{i(h-\Del'), J'}}(x_i) \right)= \nn\\
&&
-4{\hat{\alpha}_{\Del', J'}^{(\rs)} (-1)^{J'}}  {\Big \{}  \frac{\pi^\veps}{\hat{\bc}^{(\rs)}_{\tDel', \tilde{J}'}}\sum_{\sigma_t = \pm} \sin\pi\left(\tau_{\sigma_t}-\frac{1}{2}\Del_{14}^+\right)  \sin\pi\left(\tau_{\sigma_t}-\frac{1}{2}\Del_{23}^+\right)\nn\\
&&\times
\widetilde{\sum_{r, k}}  \widetilde{\sum_{ m', k'}}  
\frac{ 1}{\prod_{i=1}^4\Ga\left(\gamma^{(\rt)}_{i}-J+r+\sum_{j}k_{ji}\right)\prod_{i'=1}^4\Ga\left(\ga_{i'+}^{(\rs)}(\tDel', \tJ')-\tJ'+m'+\sum_{j'}k'_{i'j'}\right)} \nn\\
&&{\times
\sum_{(ij) = \{12, 34\}}\sum_{(i'j')=\{14, 23\}} \bbI\left[ \zeta^{(\rt\rs)}_{r, i'j'},  \tilde{\zeta}^{(\rt\rs)}_{m', ij} , \veps; 
\begin{matrix}
~\kappa_{\sigma_t}^{i(\rt)}+r, ~\kappa_{\sigma_t}^{j(\rt)}+r; ~\kappa^{i'(\rs)}_{++}(\tDel', \tJ')+m', ~\kappa^{j'(\rs)}_{++}(\tDel', \tJ')+m' \\
~1+i\sigma_t\nu,~1+\varpi_{ij}^{(\rt)}; 
~1+(\tDel-h),~1+\varpi_{i'j'}^{(\rs)}
\end{matrix} \right]}{\Big \}}\nn\\
&&-4\hat{\alpha}_{\Del', J'}^{(\rs)} (-1)^{J'}\frac{\Ga(\tJ'+2\veps)}{\Ga(\tJ'+\veps)}\frac{\Ga(\bar{\tJ}'+\veps)}{\Ga(\bar{\tJ}'+2\veps)} \left\{\tJ' \to \bar{\tJ}' = -2\veps -\tJ' \right\}
\eea
where we have again  split the contributions from $\fg_{\tDel', \tJ'}^{(\rs)}(z, \bz)$ and $\fg_{\tDel', \bar{\tJ}'}^{(\rs)}(z, \bz)$,  and have defined the integral over the cross ratios in terms of following multi-variable function:
\bea\label{Master-Int}
&&\bbI\left[\alpha, \beta, \gamma; \begin{matrix}
~a_1,~a_2;~b_1,~b_2\\
~c_1,~c_2;~d_1, d_2
\end{matrix} \right] = 
\int^1_0 dz \int^1_0 d\bz |z-\bz|^{2\gamma} (z\bz)^{\alpha} [(1-z)(1-\bz)]^{\beta} 
\tilde{\bf{F}}_4 
 \left[
\begin{matrix}
~a_1,~a_2\\
~c_1,~c_2
\end{matrix}; \ru, \rv
\right] 
\tilde{\bf{F}}_4 
 \left[
\begin{matrix}
~b_1,~b_2\\
~d_1,~d_2
\end{matrix}; \rv, \ru
\right].\nn\\
\eea
In \eqref{CrossingKernel-st-Lorentzian2}, we have also defined the following combinations of parameters for the $\fg_{\tDel', \tJ'}^{(\rs)}(z, \bz)$ contributions:
\be\label{Def:zetas}
\zeta^{(\rt\rs)}_{r, i'j'} =  \tau_{\sigma_t }-\frac{1}{4}\sum_{i=1}^4 \Del_i+r+\frac{k_{13}'+k_{24}'}{2} +\frac{\varpi_{i'j'}^{(\rs)}}{2},~~
\tilde{\zeta}^{(\rt\rs)}_{m', ij} = \tilde{\tau}'-\frac{1}{4}\sum_{i=1}^4 \bar{\Del}_i+m'+\frac{k_{12}+k_{34}}{2} +\frac{\varpi_{ij}^{(\rt)}}{2},
\ee
where  $\tilde{\tau}' = \frac{\tDel'-\tJ'}{2} = (d-1) - \tau'$ and $\tau' = \frac{\Del'-J'}{2}$ such that:
\bea\label{Def:zetas2}
\zeta^{(\rt\rs)}_{r, 14} &=& \tau_{\sigma_t}-\frac{\Del_{23}^+}{2}+k_{14}'+r, ~~ \zeta^{(\rt\rs)}_{r, 24} = \tau_{\sigma_t}-\frac{\Del_{14}^+}{2}+k_{23}'+r,\\
\tilde{\zeta}^{(\rt\rs)}_{m', 12} &=& \frac{\Del_{12}^+}{2}-{\tau}' +k_{12}+m'-1, ~~ \tilde{\zeta}^{(\rt\rs)}_{m', 34} =\frac{\Del_{34}^+}{2}-\tau' +k_{34}+m'-1.
\eea
We can similarly define the corresponding combinations for the $\fg_{\tDel', \bar{\tJ}'}^{(\rs)}(z, \bz)$ contributions by the $\tJ' \to \bar{\tJ}'$ transformation.
%%%%%%%%%%%%%%%%%%%%%%%%%%%%%%%%%%%%%%%%%
\paragraph{}
To perform the integral \eqref{Master-Int}, we first use the Mellin-Barne representation \eqref{Def:tF4} to express $\tilde{\bf F}_4$s. 
Next we perform the simpler $(z, \bz)$ integration using the Selberg integral formula
\footnote{Notice that for $d=1$, there is only one conformal cross ratio $z = \frac{|x_{12}||x_{34}|}{|x_{13}||x_{24}|}$, 
the corresponding conformal integral \eqref{Selberg2} can be further reduced to Beta-function, i. e.
\be\label{1d-ConfInt}
{\rm 1~dim.}~:~ \int^{1}_{0} dz z^{\alpha}(1-z)^{\beta} =  \frac{\Ga(\alpha+1)\Ga(\beta+1)}{\Ga(\alpha+\beta+2)} .
\ee
while for $d=2$, the integral factorizes to produce two copies of \eqref{1d-ConfInt}.}
:
\be\label{Selberg2}
\cJ(\alpha, \beta, \gamma)= \int^1_0 dz \int^1_0 d\bz (z\bz)^{\alpha}[(1-z)(1-\bz)]^{\beta}|z-\bz|^{2\gamma} = \frac{\Ga(1+\alpha, 1+\alpha+\gamma)\Ga(1+\beta, 1+\beta+\ga)\Ga(1+2\ga)}{\Ga(2+\alpha+\beta+\ga, 2+\alpha+\beta+2\ga) \Ga(1+\ga)}.
\ee
The integral \eqref{Master-Int} can now be written in terms of following Mellin-Barnes form\footnote{Here we have introduce the obvious notations: $\Ga(a_1, \dots, a_k) = \Ga(a_1)\dots \Ga(a_k)$ and 
$(a_1, \dots, a_k)_m = (a_1)_m\dots (a_k)_m$.} \footnote{It is interesting to note that the similar four variable Mell-Barnes integral was also considered in \cite{Krasnov} in computing so-called 6j-symbol for the simplest scalar exchange case. However the key difference is that the computation there was done in Euclidean signature, such that the integration range for the cross ratios $(z, \bz)$ extends the entire complex plane. In this case, the integral over $(z, \bz)$ can also be performed using Symanzik start formula after gauge fixing instead of Selberg formula, however the resultant integrand in Mellin-Barnes integral will be different.}:
\bea
&&\bbI\left[\alpha, \beta, \gamma; \begin{matrix}
~a_1,~a_2;~b_1,~b_2\\
~c_1,~c_2;~d_1, d_2
\end{matrix} \right] = \frac{\pi^4 } {\sin\pi c_1 \sin\pi c_2 \sin\pi d_1 \sin\pi d_2} 
\int^{i\infty}_{-i\infty} \frac{dx}{2\pi i} \int^{i\infty}_{-i\infty} \frac{dy}{2\pi i} \int^{i\infty}_{-i\infty} \frac{dx'}{2\pi i} \int^{i\infty}_{-i\infty} \frac{dy'}{2\pi i} 
(-1)^{s+t+s'+t'}\nn\\
&&\times
{\Ga(-x)\Ga(-y)\Ga(-x')\Ga(-y')}\frac{\Ga(a_1+x+y, a_2+x+y, b_1+x'+y', b_2+x'+y')}{\Ga(c_1+x, c_2+y, d_1+x', d_2+y')} \cJ(\alpha+x+y', \beta+y+x', \ga)\nn\\
&&=
\frac{\pi^4 }{\sin\pi c_1\sin\pi c_2 \sin\pi d_1 \sin\pi d_2}
\Ga(a_1, a_2)\Ga(b_1, b_2)\cJ(\alpha, \beta, \gamma)\nn\\
&&\times
\sum_{m_1, m_2, n_1, n_2 =0}^{\infty}\frac{1}{m_1! n_1! m_2! n_2! }\frac{(a_1, a_2)_{m_1+n_1}(b_1, b_2)_{m_2+n_2}}{ \Ga(c_1+m_1, c_2+n_1,d_1+n_2, d_2+m_2)} \frac{ \cJ(\alpha+m_1+m_2, \beta+n_1+n_2, \ga)} {\cJ(\alpha, \beta, \ga)}.
\label{Master-Int0}
\eea
In evaluating \eqref{Master-Int0}, we have enclosed the poles at $s, t, s', t' = {\mathbb{Z}}_{\ge 0}$, and express the resultant infinite summations over ratios only containing Pochhammer symbols in the numerator or $\Ga$-function in the denominator, i.e. the parameters involved in the summation do not introduce additional singularities.
It would be interesting to know if we can express this infinite summation in terms of a possible four variable generalization of hypergeometric function, 
similar to the ones considered in \cite{Exton-Four}.
To the best of our knowledge however,
we can express the infinite series in \eqref{Master-Int0} into a more compact form as the following double integral:
\bea\label{Master-Int00}
&&\bbI\left[\alpha, \beta, \gamma; \begin{matrix}
~a_1,~a_2;~b_1,~b_2\\
~c_1,~c_2;~d_1, d_2
\end{matrix} \right]  \nn\\
&&
= \frac{\pi^4}{\sin\pi c_1\sin\pi c_2 \sin\pi d_1 \sin\pi d_2} \frac{\Ga(1+2\ga)}{\Ga(1+\ga)} \int^{i\infty}_{-i\infty} \frac{dx}{2\pi i}\int^{i\infty}_{-i\infty} \frac{dy}{2\pi i}
\Xi\left[\beta, x; \begin{matrix}
~a_1,~a_2\\
~c_2
\end{matrix} \right] 
\Xi\left[\alpha, y; \begin{matrix}
~b_1,~b_2\\
~d_2
\end{matrix} \right]\nn\\
&&
\times \footnotesize{\tilde{\bbF}^{0, 4}_{2, 1}\left[ \begin{matrix}
~\cdotp:~a_1+x, a_2+x,   1+\beta+y , 1+\beta+\ga+y; b_1+y, b_2+y, 1+\alpha+x, 1+\alpha+\ga+x \\
~2+\alpha+\beta+\ga+x+y, 2+\alpha+\beta+2\ga+x+y :~c_1 ; d_1
\end{matrix}; 1, 1\right] }.
\eea
Here the measure factor is:
\be\label{Def:Xi-measure}
\Xi\left[\beta, x; \begin{matrix}
~a_1,~a_2\\
~c_2
\end{matrix} \right] = (-1)^x \frac{\Ga(-x)\Ga(a_1+x, a_2+x)\Ga(1+\beta+x, 1+\beta+\ga+x)}{\Ga(c_2+x)},
\ee
the second measure factor in \eqref{Master-Int00} is given by obvious exchange of parameters $(\beta, x, a_1, a_2, c_2) \to (\alpha, y, b_1, b_2, d_2)$.
The integration contours for $(x, y)$ again close in the right half plane to pick up the poles of $\Ga(-x)$ and $\Ga(-y)$ at $x, y = 0, 1, 2, \dots$.
Here we have introduced Kamp\'{e} de F\'{e}riet hypergeometric function which is a further two variable generalization of hypergeometric function:
\be\label{Def:KdF function}
{\bbF}^{p, q}_{r, s}\left[ \begin{matrix}
~a_1,\dots, a_p:~b_1, \dots, b_q; b_1', \dots, b_q' \\
~c_1,\dots, c_r:~d_1, \dots, d_s; d_1', \dots, d_s'
\end{matrix}; x, y\right] 
= \sum_{m\ge 0} \sum_{n\ge 0}\frac{(a_1, \dots , a_p)_{m+n}}{(c_1, \dots, c_r)_{m+n}}\frac{(b_1, \dots , b_q)_m (b_1', \dots ,b_q')_n}{ (d_1, \dots ,d_s)_m (d_1', \dots , d_s')_n}\frac{x^m y^n}{m! n!}\,,
\ee
and its regularized version:
\bea\label{Def:tKdF function}
&&\tilde{\bbF}^{p, q}_{r, s}\left[ \begin{matrix}
~a_1,\dots, a_p:~b_1, \dots, b_q; b_1', \dots, b_q' \\
~c_1,\dots, c_r:~d_1, \dots, d_s; d_1', \dots, d_s'
\end{matrix}; x, y\right] \nn\\
&&=\frac{1}{\prod^{r}_{k=1}\Ga(c_k)\prod_{l=1}^s \Ga(d_l)\Ga(d_l')} {\bbF}^{p, q}_{r, s}\left[ \begin{matrix}
~a_1,\dots, a_p:~b_1, \dots, b_q; b_1', \dots, b_q' \\
~c_1,\dots, c_r:~d_1, \dots, d_s; d_1', \dots, d_s'
\end{matrix}; x, y\right],
\eea
which does not contain any singularities in all the parameters $\{a_1, \dots a_p\}$, $\{c_1, \dots, c_r\}$ and others, similar to regularized Gauss hypergeometric function.
In other words we can express \eqref{Master-Int0} in terms of a double infinite summation of regularized Kamp\'{e} de F\'{e}riet functions of unit arguments\footnote{Very recently, Kamp\'{e} de F\'{e}riet function has also appeared in the computation of crossing kernel in the simplified light cone and identical scalar limit \cite{Li}, as it naturally arises from the inner product of two single variable hypergeometric functions.}.
\paragraph{}
Despite the complicated looking expression of crossing kernel in \eqref{CrossingKernel-st-Lorentzian2}, we can however use the
expression in \eqref{Master-Int0} and \eqref{Master-Int00} to extract the important information such as the operator spectrum.
More precisely, consider the dependence of $(\zeta^{(\rt\rs)}_{r, i'j'}, \tilde{\zeta}^{(\rt\rs)}_{m', ij})$ defined in \eqref{Def:zetas} and \eqref{Def:zetas2}, 
all the potential singularities associated with them are contained in the $\Gamma$-functions of the measure factors \eqref{Def:Xi-measure}.
The remaining dependences of $(\zeta^{(\rt\rs)}_{r, i'j'}, \tilde{\zeta}^{(\rt\rs)}_{m', ij})$ are contained in the parameters of the regularized Kamp\'{e} de F\'{e}riet functions, such that they only play the role of residues in the analysis.
%\paragraph{}
Let us focus on $\fg_{\tDel', \tJ'}^{(\rs)}(z, \bz)$ contributions in \eqref{CrossingKernel-st-Lorentzian2}, and their singularities appear through the overall $\Ga$-functions in \eqref{Master-Int0}:
\be\label{singular1}
\Ga\left(1+\tilde{\zeta}^{(\rt\rs)}_{m', ij}\right) \Ga\left(1+\tilde{\zeta}^{(\rt\rs)}_{m', ij}+\veps\right),~~ ij =12, 34,
\ee 
while keeping fixed t-channel $(\nu, J)$, they encode the spectrum of s-channel exchange operators when expanding t-channel conformal partial waves in terms of s-channel ones.
For arbitrary $\veps$, the simple poles come from the first $\Ga$-function in \eqref{singular1} and are located at:
\bea\label{CK-poles1}
1+\tilde{\zeta}^{(\rt\rs)}_{m', 12} &=& \frac{\Del_{12}^+}{2}-\frac{(\Del'-J')}{2} +k_{12}+m'= 0, -1, -2, \dots,\\
1+ \tilde{\zeta}^{(\rt\rs)}_{m', 34}&=& \frac{\Del_{34}^+}{2}-\frac{(\Del'-J')}{2} +k_{34}+m'  = 0, -1, -2, \dots,
\eea
they precisely correspond to the following two infinite towers of double trace operators:
\bea\label{CK-poles-a}
\Del'-J'= \Del_1+\Del_2+2n_{12}, \quad \Del'-J'= \Del_3+\Del_4+2n_{34}, ~~n_{12}, n_{34}\in {\mathbb{Z}}_{\ge 0}.
\eea
Similarly, when we now consider the singularities from $\fg_{\tDel', \bar{\tJ}'}^{(\rs)}(z, \bz)$ contributions, the singularities now located at:
\bea\label{CK-poles2}
1+\bar{\tilde{\zeta}}^{(\rt\rs)}_{m', 12} &=& \frac{\Del_{12}^+}{2}-\frac{(d-\Del' - J')}{2} +k_{12}+m'= 0, -1, -2, \dots,\\
1+ \bar{\tilde{\zeta}}^{(\rt\rs)}_{m', 34}&=& \frac{\Del_{34}^+}{2}-\frac{(d-\Del'-J')}{2} +k_{34}+m'  = 0, -1, -2, \dots,
\eea
these correspond to the shadows of the two infinite towers in \eqref{CK-poles-a}:
\bea\label{CK-poles-b}
(d-\Del')-J'= \Del_1+\Del_2+2\bar{n}_{12}, \quad (d-\Del')-J'= \Del_3+\Del_4+2\bar{n}_{34}, ~~\bar{n}_{12}, \bar{n}_{34}\in {\mathbb{Z}}_{\ge 0}.
\eea
The double trace poles \eqref{CK-poles-a} and \eqref{CK-poles-b} are consistent with the general comment about the singularity structure in the crossing kernel made in \cite{Liu-2018},
it would be interesting to perform the full analysis of the singularity structures of the crossing kernel \eqref{CrossingKernel-st-Lorentzian2} and the explicit residues in terms of hypergeometric functions with multiple variables, as they correspond to the various expansion coefficients.

\section{Orthogonality and Decomposition in AdS Space}\label{Sec:AdS}
\paragraph{}
In this section, we present few complementary computations using the holographic dual configuration of the conformal partial wave \eqref{Psi-s-full} to demonstrate they indeed form an orthogonal basis in $d+1$-dimensional Euclidean Anti-de Sitter space. In particular, as an advantage of this approach, in demonstrating the orthogonality we do not need the explicit integrated form of kinematical basis but only the property of Euclidean AdS harmonic function.
We will next demonstrate how contact Witten diagrams can be decomposed in terms of them.

%%%%%%%%%%%%%%%%%%%%%%%%%%%%%%%%%%%%%%%%
\subsection{Conformal Partial Wave in AdS Space}
\paragraph{}
%%%%%%%%%%%%%%%%%%
{
We start with the following definition of CPW written in the embedding space coordinates where we will be mostly working in this section:
\be\label{Def:psi}
\Psi^{(\rs)}_{\nu,J}(P_i) = 
\frac{1}{J!(h-1)_J \pi^h}\int_{\partial {\text{AdS}}_{d+1}} d^dP_0 
\langle
\cO_{\Delta_1}(P_1)
\cO_{\Delta_2}(P_2)
\cO_{h+i\nu,J}(P_0,\cD_{Z_0})
\rangle
\langle
\tilde{\cO}_{h-i\nu,J}(P_0,Z_0)
\cO_{\Delta_3}(P_3)
\cO_{\Delta_4}(P_4)
\rangle,\\
\ee
for a good review on embedding space and our conventions, please refer to \cite{CKK:2017}.
%%%%%%%%%%%%%%%%%%%
\begin{figure}[h]
\centering
\includegraphics[scale=0.15]{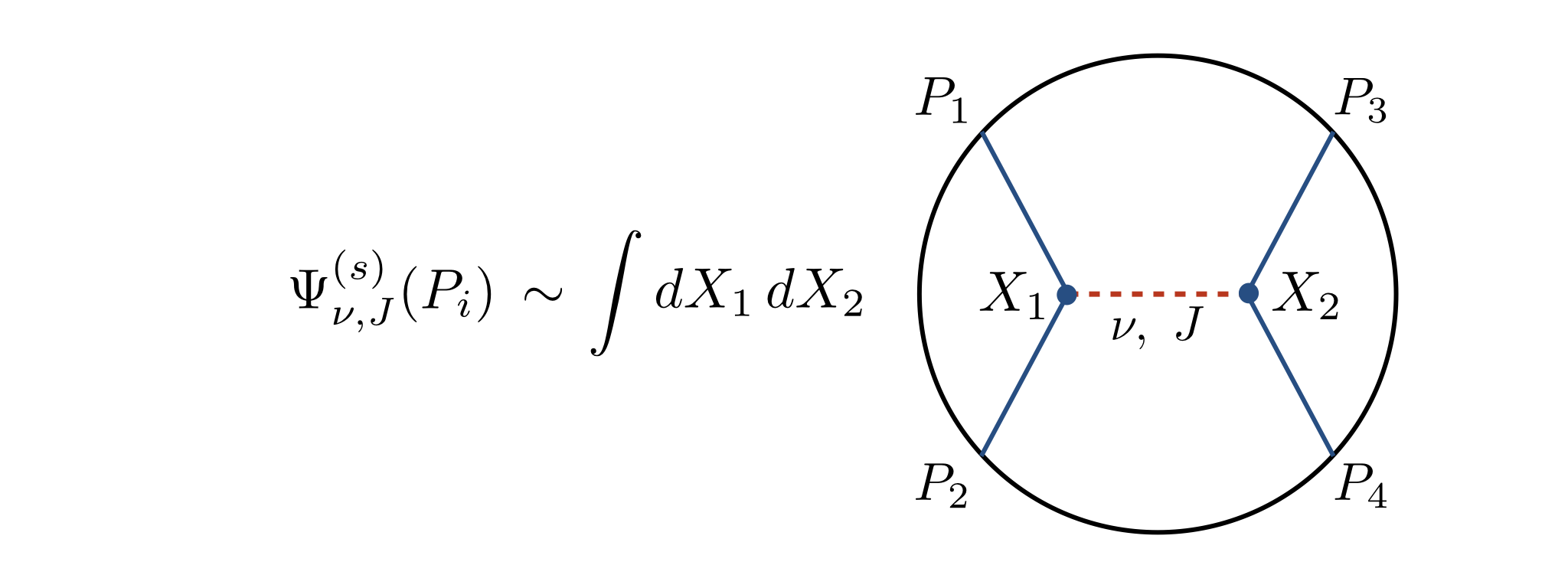}
\caption{$\Psi_{\nu, J}^{(\rs)}(P_i)$ can be expressed as a bulk exchange diagram. 
Here the blue lines are bulk-to-boundary propagators and the internal dashed line denotes the AdS harmonic function.}
\label{3: fig bulk-CPW}
\end{figure}
%%%%%%%%%%%%%%%%%%
As the three point CFT correlation function and three point Witten diagram are proportional up to an dynamical factor, the definition of conformal partial wave can be expressed in terms of a integral over d+1 dimensional AdS space \cite{SpinningAdS}:
\bea\label{3: bulk-cpw1}
\Psi^{(\rs)}_{\nu,J}(P_i)
&=&
\frac{1}{\cB^{\Delta_1,\Delta_2}_{h+i\nu,J}\cB^{\Delta_3,\Delta_4}_{h-i\nu,J}}
\frac{1}{J! (h-1)_J \pi^h}\left[\frac{1}{J! (h-\frac{1}{2})_J }\right]^2
\int_{\partial {\text{AdS}}_{d+1}} dP_0\\
&&\times 
\int_{\text{AdS}_{d+1}} dX_1 ~ \Pi_{\Delta_1}(X_1,P_1)\, (K_1\cdot \nabla_1)^J\Pi_{\Delta_2}(X_1,P_2)~\Pi_{h+i\nu,J}(X_1,P_0;W_1,\cD_{Z_0})
\nn\\
&&\times 
\int_{\text{AdS}_{d+1}} dX_2~ \Pi_{\Delta_3}(X_2,P_3)\, (K_2\cdot \nabla_2)^J\Pi_{\Delta_4}(X_2,P_4)~\Pi_{h-i\nu,J}(X_2,P_0;W_2,Z_0)\nn\,.
\eea
where $\cB^{\Delta_1,\Delta_2}_{h+i\nu,J}$ and $\cB^{\Delta_3,\Delta_4}_{h-i\nu,J}$ are the dynamical factors arising from integrating the interaction vertices over the entire AdS space.
The bulk to boundary propagator for spin $J$ tensor field is given by: 
\be\label{AdS-Propagator-J}
\Pi_{\Del, J}(X, P) =\cC_{\Del, J}\frac{((-2P\cdot X)(W\cdot Z)-(2W\cdot P)(Z\cdot X))^J}{(-2P\cdot X)^{\Del+J}},
\ee
where $\{P_i, Z_i\}$ denote the boundary position and polarization vectors, and $\{X_i, W_i\}$ denote their AdS counterparts. 
The dynamical factor is fixed through the following integral:
\bea
&&\frac{1}{J!\left(h-\frac{1}{2}\right)_J}
\int_{\text{AdS}_{d+1}} dX~
\Pi_{\Delta_1}(X,P_1)\,
(K\cdot\nabla)^J
\Pi_{\Delta_2}(X,P_2)\,
\Pi_{\Delta_0}(X,P_0;W,{Z_0})\\[5pt]
&& \qquad \qquad\qquad
= \cB^{\Delta_1,\Delta_2}_{\Delta_0;J}~
\langle
\cO_{\Delta_1}(P_1)
\cO_{\Delta_2}(P_2)
\cO_{\Delta_0,J}(P_0,{Z_0})
\rangle,\nn
\eea
Explicitly $\cB^{\Delta_1,\Delta_2}_{\Del
_0,J}$ is given by:
\be
\cB^{\Delta_1,\Delta_2}_{\Del_0 ,J}
=
\frac{\pi^h}{2}(-2)^J\cC_{\Delta_1}\cC_{\Delta_2}\cC_{\Delta_0,J} 
\times 
\frac{\Gamma\left(\frac{\sum_{i=0}^2\Delta_i+J-d}{2}\right)}{\Gamma(\Delta_1)\Gamma(\Delta_2)\Gamma(\Delta_0+J)}
\Gamma\left(\frac{\Delta_0\pm\Delta_{12}^- +J}{2}\right)\Gamma\left(\frac{\Delta_{12}^+-\Delta_0+J}{2}\right)\,.
\ee
In this expression, the $P_0$ integration is nothing but the definition of the AdS harmonic function.
By substituting its definition:
\be\label{Def:AdS-Harmonic}
\Omega_{\nu, J}(X_1, X_2; W_1, W_2) =\frac{\nu^2}{\pi J! (h-1)_J} \int_{\rm{\partial AdS_{d+1}}} dP_0 \Pi_{h+i\nu, J} (X_1, P_0; W_1, D_Z ) \Pi_{h-i\nu, J}(X_2, P_0; W_2, Z),
\ee
we obtain the AdS representation of the conformal partial wave \eqref{Psi-s-full} in terms of AdS harmonic function:
\bea\label{3: bulk-cpw}
\Psi^{(\rs)}_{\nu,J}(P_i)
&=&
\frac{1}{\pi^h}\frac{\pi}{\nu^2~\cB^{\Delta_1,\Delta_2}_{h+i\nu,J}\cB^{\Delta_3,\Delta_4}_{h-i\nu,J}}
\left[\frac{1}{J! (h-\frac{1}{2})_J}\right]^2
\int_{\text{AdS}_{d+1}} \! dX_1\,dX_2 ~ \Pi_{\Delta_1}(X_1,P_1)~ (K_1\cdot \nabla_1)^J\Pi_{\Delta_2}(X_1,P_2)~
\nn\\
&\times& 
 \Pi_{\Delta_3}(X_2,P_3)~ (K_2\cdot \nabla_2)^J\Pi_{\Delta_4}(X_2,P_4)~\Omega_{\nu,J}(X_1,X_2;W_1,W_2)\,.
\eea
Diagrammatically, this expression can be described as in Fig. \ref{3: fig bulk-CPW}\,.
As an important property of the AdS harmonic function, it satisfies the following orthogonality relation:
\bea\label{orthogonal omega}
&&\frac{1}{J!\left(h-\frac{1}{2}\right)_J}\int_{\text{AdS}_{d+1}} dX_0 ~
\Omega_{\nu,J}(X_1,X_0;W_1,K_0)\, \Omega_{\nu',J'}(X_0,X_2;W_0,W_2)\nn\\[8pt]
&&
=
\frac{1}{2}\,\delta_{J,J'} \,\left[\delta(\nu-\nu')+\delta(\nu+\nu')\right]\, 
\Omega_{\nu,J}(X_1,X_2;W_1,W_2)\,.
\eea
Let us briefly comment here on the connections between our computation in this section, and another well-known 
holographic dual configuration of conformal partial waves in the literature, i. e.  ``geodesic Witten diagram'' in \cite{GWD1}.
As explicitly demonstrated in \cite{CKK:2017} using the split representation, when we decompose the four point  geodesic Witten diagram into its kinematic building blocks in this construction, which was called ``three point geodesic Witten diagram'' and was in turn proportional to the three normal Witten diagram, the four Witten exchange diagram is proportional to conformal partial wave up to an overall dynamical factor\footnote{Please note however the overall dynamical factor depends on spectral parameter, such that upon integration yields combination of single and double trace operators.}.
This relation therefore also allows us to express conformal partial wave $\Psi_{\nu, J}^{(\rs)}(x_i)$ in \eqref{Psi-s-full} in terms of the AdS bulk integral \eqref{3: bulk-cpw1}.
%%%%%%%%%%%%%%%%%%%%%%%%%%%%%
 %%%%%%%%%%%%%%%%%%%%%%%%%%%%%
\subsection{Orthogonality in AdS space}
\paragraph{}
Having rewritten the conformal partial wave in terms of AdS integral \eqref{Psi-s-full}, we will see their orthogonality can be identified directly as the consequence of orthogonality of the Euclidean AdS harmonic function.
Let us reconsider the inner product of two ${ \Psi}_{\nu, J}^{(\rs)}(P_i)$s which appears in the LHS of the orthogonality relation in terms of these AdS integral\footnote{Here in writing out the inner product, we drop the dependence of the coordinate such $P_i$ on LHS which are integrated over.}:
\bea\label{3: psi-stos}
\left({\Psi}^{(\rs)}_{\nu,J},{ \Psi}^{(\rs)}_{\nu',J'}\right)
=\int_{\partial {\text{AdS}}_{d+1}} \frac{\prod_{i=1}^4 d^dP_i}{\text{vol}\left(SO(1,d+1)\right)}~
\Psi^{(\rs)}_{\nu,J}(P_i)\,\overline{\Psi}^{(\rs)}_{\nu',J'}(P_i).
\eea
Here for the two $\Psi_{\nu, J}^{(\rs)}$s, by substituting the bulk integral form \eqref{3: bulk-cpw}, the inner product \eqref{3: psi-stos} can be represented as follows:
\bea \label{3: IP-CPW comp1}
&&
\left(\Psi^{(\rs)}_{\nu,J},{\Psi}^{(\rs)}_{\nu',J'}\right)
=\frac{1}{\pi^{2h}}
\frac{1}
{\cB^{\Delta_1,\Delta_2}_{h+i\nu,J}\cB^{\Delta_3,\Delta_4}_{h-i\nu,J}\,
\cB^{d-\Delta_1,d-\Delta_2}_{h-i\nu',J'}\cB^{d-\Delta_3,d-\Delta_4}_{h+i\nu',J'}}
\frac{\pi^2}{\nu^2 ~\nu'^2}\left[\frac{1}{J!\,(h-\frac{1}{2})_J}\right]^4 
\nn \\
&&\times 
\int_{\mathbb{R}^d} \frac{\prod d^d P_i}{\text{Vol}\left(SO(1,d+1)\right)}
\int_{\text{AdS}_{d+1}} dX^{12}dX^{34}d\tilde{X}^{12}d\tilde{X}^{34}\nn\\
&&\times
\Pi_{\Delta_1}(P_1, X^{12})\,\Pi_{d-\Delta_1}(P_1, \tilde{X}^{12})\,
(K^{12}\cdot \nabla_{12})^J\Pi_{\Delta_2}(P_2, X^{12})\,(\tilde{K}^{12}\cdot \tilde{\nabla}_{12})^J\Pi_{d-\Delta_2}(P_2, \tilde{X}^{12})\nn\\
&&\times
\Pi_{\Delta_3}(P_3, X^{34})\,\Pi_{d-\Delta_3}(P_3, \tilde{X}^{34})\,
(K^{34}\cdot \nabla_{34})^J\Pi_{\Delta_4}(P_4, X^{34})\,(\tilde{K}^{34}\cdot \tilde{\nabla}_{34})^J\Pi_{d-\Delta_4}(P_4, \tilde{X}^{34})\nn\\
&&\times \Omega_{\nu,J}(X^{12},\tilde{X}^{12} ; W^{12}, \tilde{W}^{12})~ \Omega_{\nu,J}(X^{34},\tilde{X}^{34} ; W^{34}, \tilde{W}^{34}).
\eea
%%%%%%%%%%%%%%%
\begin{figure}[h]
\begin{center}
\includegraphics[height=6cm]{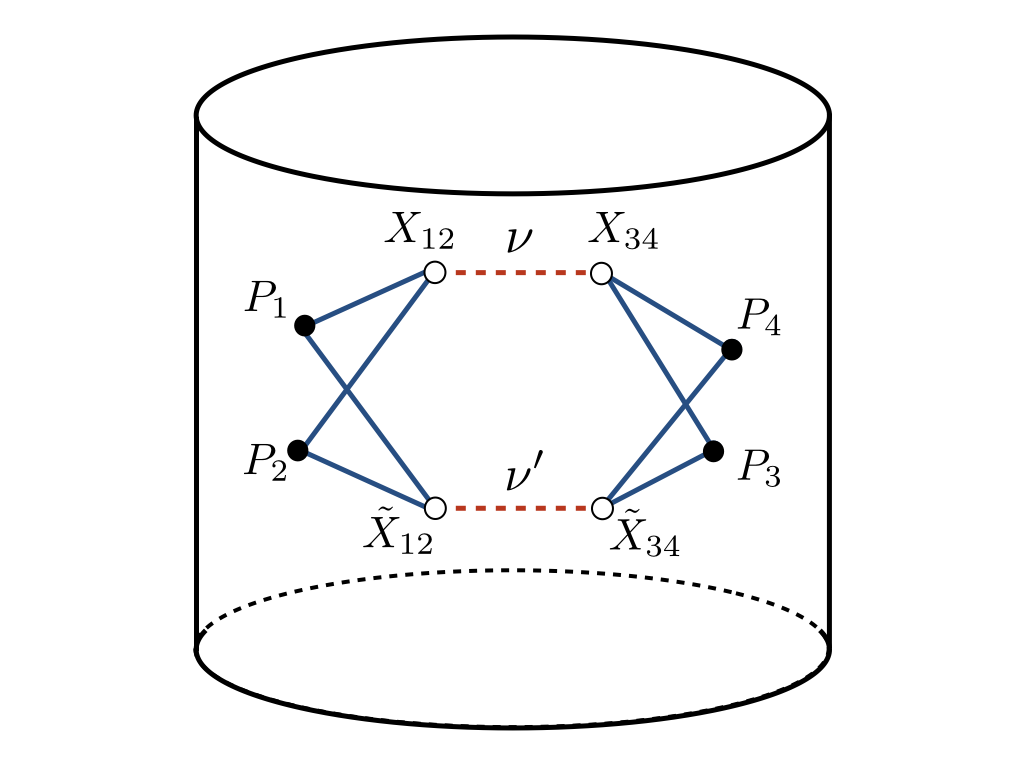}
\caption{\label{3: Fig:stos}
The diagrammatic expression of \eqref{3: IP-CPW comp1}. The white circles are bulk points and the black circles are boundary points.}
\end{center}
\end{figure}
%%%%%%%%%%%%%%%%
Here $X^{12}$, $X^{34}$, $\tilde{X}^{12}$ and $\tilde{X}^{34}$ are bulk interaction points to be integrated over AdS,
the indices in $X^{ij}$ and $\tilde{X}^{ij}$ denote the specific boundary points it is connected with.
In each bulk point $X^{ij}$ and $\tilde{X}^{ij}$, we have chosen particular interactions with AdS covariant derivatives.
Although we can choose another type of interaction vertices, the final result would not be changed. 
In the above expression, each boundary integral has the same form of the definition of the AdS harmonic function \eqref{Def:AdS-Harmonic} again, 
and the pairs of bulk-to-boundary propagators are also combined into AdS harmonic functions.
Finally, the inner product becomes the following bulk integral:
\bea\label{3: stos-bulk}
\left(\Psi^{(\rs)}_{\nu,J},{\Psi}^{(\rs)}_{\nu',J'}\right)&=&
\cN^{\Delta_i}_{\nu,\nu';J,J'} \left[\frac{1}{J!\,(h-\frac{1}{2})_J}\right]^4
\int_{AdS_{d+1}} 
\frac{dX^{12}dX^{34}d\tilde{X}^{12}d\tilde{X}^{34}}
{\text{Vol}\left(SO(1,d+1)\right)}\nn\\
&&\times
\Omega_{\alpha_1}(X^{12},\tilde{X}^{12})
(\tilde{K}^{12}\cdot \tilde{\nabla}_{12})^J
(K^{12} \cdot \nabla_{12})^J
\Omega_{\alpha_2}(X^{12},\tilde{X}^{12})\nn\\
&&\times
\Omega_{\alpha_3}(X^{34},\tilde{X}^{34})
(\tilde{K}^{34}\cdot \tilde{\nabla}_{34})^J
(K^{34}\cdot \nabla_{34})^J
\Omega_{\alpha_4}(X^{34},\tilde{X}^{34})\nn\\
&&\times
\Omega_{\nu,J}(X^{12},X^{34};W^{12},W^{34})
\Omega_{\nu',J}(\tilde{X}^{12},\tilde{X}^{34};\tilde{W}^{12},\tilde{W}^{34})\delta_{J, J'}.
\eea
Here in the indices of harmonic functions, we have introduced the associated spectral parameters $\{\alpha_i\}$ through the relation $\Delta_i=h+i \alpha_i$,
and clearly for $J \neq J'$ the contraction of polarization vectors vanish.
The coefficients are combined as $\cN^{\Delta_i}_{\nu,\nu';J,J'}$ which is given as:
\bea
\cN^{\Delta_i}_{\nu,\nu';J,J'}=\frac{1}{\pi^{2h}}
\frac{1}
{\cB^{\Delta_1,\Delta_2}_{h+i\nu,J}\cB^{\Delta_3,\Delta_4}_{h-i\nu,J}\,
\cB^{d-\Delta_1,d-\Delta_2}_{h-i\nu',J'}\cB^{d-\Delta_3,d-\Delta_4}_{h+i\nu',J'}}
\left(\prod_{i=1}^4\frac{\pi}{\alpha_i^2}\right)\frac{\pi^2}{\nu^2 \, \nu'^2}\delta_{J, J'}\,.
\eea
Diagrammatically, the inner product can be expressed as in Fig.\ref{3: Fig:stos}.
Each dashed line in the bulk is an AdS harmonic function, not  the usual AdS bulk to bulk propagator. 
The blue dashed lines are the scalar functions and the red dashed lines are the functions with spin.
%%%%%%%%%%%%%%%%%%%%%%%%%%%%%%
\begin{figure}[h]
\begin{center}
\includegraphics[height=5cm]{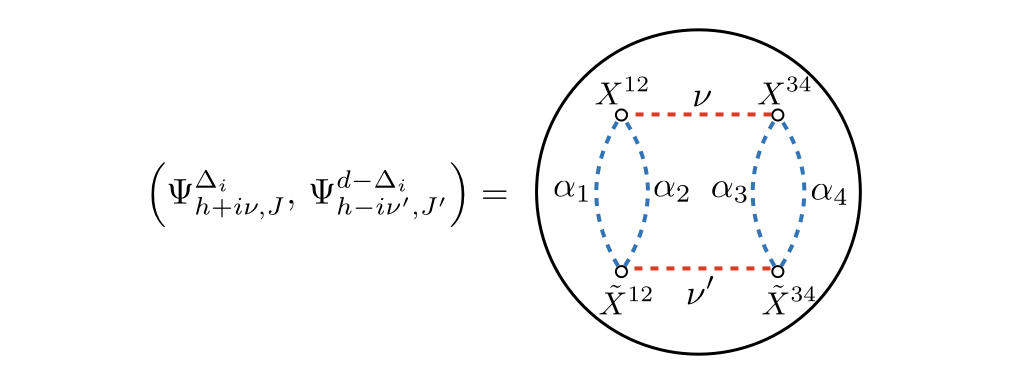}
\caption{\label{3: Fig:stos}
The inner product of two ${\Psi}_{\nu, J}(P_i)$s can interpreted as a bubble diagram in bulk. The bulk points are integrated over AdS. }
\end{center}\end{figure}
%%%%%%%%%%%%%%%%%%%%%%%%%%%%%%
To compute the diagram in Fig.\ref{3: Fig:stos}, we need to evaluate the following bulk integrals with three harmonic functions.
\bea
&&\Xi^{\alpha_1,\alpha_2}_{\nu, J} (X_1,X_2;W_1,W_2) =\label{3: Def:Xi}\\
&&\frac{1}{J! (h-\frac{1}{2})_J} \int_{\text{AdS}_{d+1}} dY~
\Omega_{\alpha_1} (X_1,Y)  \,
(W_1\cdot \nabla_1)^J (K_Y\cdot\nabla_Y)^J
\Omega_{\alpha_2}(X_1,Y)\,
\Omega_{\nu,J} (Y,X_2;W_Y,W_2)  \,.\nn
\eea
Each bulk point in Fig. \ref{3: Fig:stos} has the same form as $\Xi_{\nu, J}^{\alpha_1, \alpha_2}(X_1, X_2; W_1, W_2)$, 
and it is the building block of what we call the ``bubble diagram''.
The explicit computation of $\Xi_{\nu, J}^{\alpha_1, \alpha_2}(X_1, X_2; W_1, W_2)$ is done in Appendix \ref{Appendix:Xi}.
%%%%%%%%%%%%%%%%%%%%%%%%%%%%
\begin{figure}[h]
\begin{center}
\includegraphics[scale=0.30]{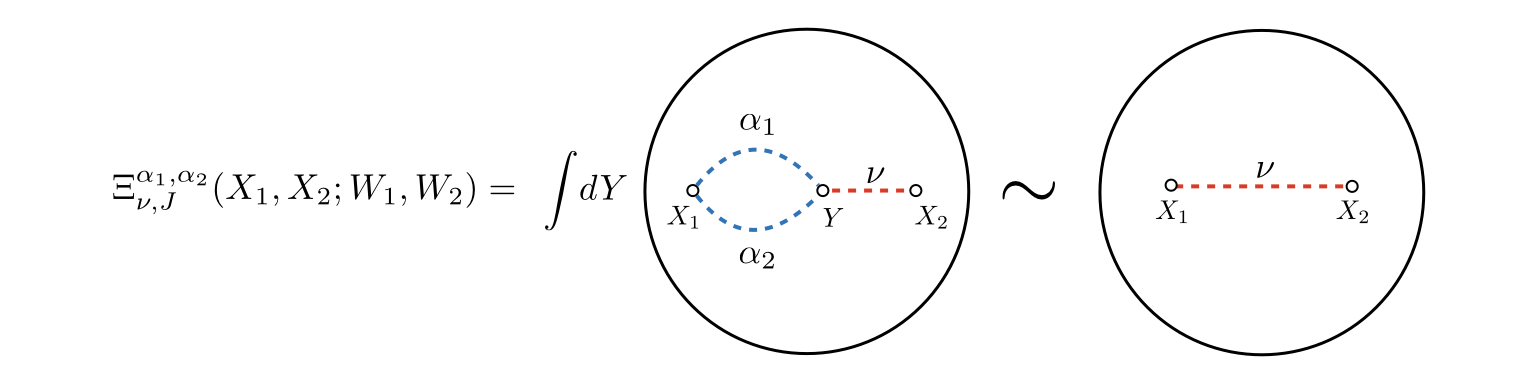}
\caption{\label{3: fig Comp-Xi}
The summary of the calculation of $\Xi_{\nu, J}^{\alpha_1, \alpha_2}(X_1, X_2; W_1, W_2)$. According to the completeness of the AdS harmonic functions, 
a loop of harmonic functions can be expanded as a series of harmonic functions and due to the orthogonality, finally, 
it is proportional to a single harmonic function. }
\end{center}\end{figure}
%%%%%%%%%%%%%%%%%%%%%%%%%%%%%%%
Applying the above result to $X^{12}$ and $\tilde{X}^{34}$ integral in \eqref{3: stos-bulk}, the inner product can be simplified as:
\bea
\left(\Psi^{(\rs)}_{\nu,J}, {\Psi}^{(\rs)}_{\nu',J'}\right)
&=&
\cN^{\Delta_i}_{\nu,\nu';J, J'} \left[\frac{1}{J!\,(h-\frac{1}{2})_J}\right]^2
\int_{\text{AdS}_{d+1}} \frac{dX^{34}d\tilde{X}^{12}}{\text{vol}\left(SO(1,d+1)\right)}\\
%[5pt]
&\times& 
F(\alpha_1,\alpha_2,\nu)\, \Omega_{\nu,J} (\tilde{X}^{12},X^{34};\tilde{K}^{12},W^{34})\,
F(\alpha_3,\alpha_4,\nu')\, \Omega_{\nu',J} (\tilde{X}^{12},X^{34};\tilde{W}^{12},\tilde{K}^{34})\delta_{J,J'}\,.\nn
\eea
Now we can use the orthogonality of the AdS harmonic function which is given in \eqref{orthogonal omega} for one of the bulk integrals
to conclude that the inner product of two $\Psi_{\nu, J}$s is given in the following form:
\bea
\left(\Psi^{(\rs)}_{\nu,J},{\Psi}^{(\rs)}_{\nu',J'}\right)
&=&
\cN^{\Delta_i}_{\nu,\nu';J, J'}\, 
F(\alpha_1,\alpha_2,\nu)F(\alpha_3,\alpha_4,\nu')
\frac{1}{2}\left[\delta(\nu-\nu')+\delta(\nu+\nu')\right]\del_{J, J'}\nn\\
&&\times 
\frac{1}{J!\,(h-\frac{1}{2})_J}
\Omega_{\nu,J} (X,X;K,W)
\int_{\text{AdS}_{d+1}} \frac{dX}{\text{vol}\left(SO(1,d+1)\right)}\nn\\
&=&\frac{1}{2}\, n_{\nu,J} \left[\delta(\nu-\nu') +\cK_{\nu}\, \delta(\nu+\nu') \right]\delta_{J, J'},
\eea
where $\cK_{\nu}$ is given by:
\be \label{coef: K}
\cK_{\nu}=\frac{\Gamma\left(\ow_- \pm \ra^{(\rs)} \right)\Gamma\left(\ow_+ \pm \rb^{(\rs)}\right)}
{\Gamma\left(\ow_+\pm \ra^{(\rs)} \right)\Gamma\left(\ow_- \pm \rb^{(\rs)} \right)}\,,
\ee
and this factor comes from the ratio of $\cN_{\nu,\nu';J, J'}^{\Delta_i}$ and $F(\alpha_i,\alpha_j,\nu^{(\prime)})$.
$\Omega(X, X; K, W)$ is the norm of the AdS harmonic function \eqref{Def:AdS-Harmonic} by setting $X=Y$ and using $X^2=-1$.
The normalization factor $n_{\nu,J}$ is now given as:
\bea
n_{\nu,J}&=&
\cN^{\Delta_i}_{\nu,\nu;J, J}\, 
F(\alpha_1,\alpha_2,\nu)F(\alpha_3,\alpha_4,\nu)
\frac{1}{J!\,(h-\frac{1}{2})_J}
\Omega_{\nu,J} (X,X;K,W)
\int_{\text{AdS}_{d+1}} \frac{dX}{\text{Vol}\left(SO(1,d+1)\right)}\nn\\
&=&
\left(
\frac{\pi}{\nu^2}\frac{\Gamma(J+1)}{2^{J-1}\, \Gamma(h+J)}
\right)^2
\frac{1}{J!\,(h-\frac{1}{2})_J}
\frac{\Omega_{\nu,J} (X,X;K,W)}{(\cC_{h\pm i\nu,J})^2}
\int_{\text{AdS}_{d+1}} \frac{dX}{\text{Vol}\left(SO(1,d+1)\right)}.
\eea 
Here we can evaluate the normalization factor of the AdS harmonic function as:
\bea
\frac{1}{J!\,(h-\frac{1}{2})_J}
\frac{\Omega_{\nu,J} (X,X;K,W)}{(\cC_{h\pm i\nu,J})^2}
= 
\frac{\pi^h\Gamma(2h+J)\Gamma(h)}{\Gamma(J+1)\Gamma(2h)^2} \,\frac{\nu^2}{\pi}\frac{1}{\cC_{h\pm i\nu,J}}\,,
\eea 
and the bulk integration is evaluated as well:
\bea
\int_{\text{AdS}_{d+1}} dX=\text{Vol}(\text{AdS}_{d+1})=\frac{\text{Vol}\left(SO(1,d+1)\right)}{\text{Vol}\left(SO(d+1)\right)}\,.
\eea 
The volume of $SO(1,d+1)$ is infinite because it is a non-compact group, however, 
this factor is precisely cancelled by the regularization factor in the definition of the inner product.
\paragraph{}
In the following sections, using this expression, we will consider conformal block decompositions of certain AdS Witten contact diagram as an example.
By applying the Euclidean inversion formula, an arbitrary four-point AdS contact Witten diagram $\cA(P_i)$ can also be decomposed by CPWs as follows:
\bea
\cA(P_i)= 
\sum_{J=0}^\infty \int_{-\infty}^\infty \frac{d\nu}{n_{\nu,J}} ~ 
\left(\cA, {\Psi}^{(\rs)}_{\nu,J}\right) \, 
\Psi^{(\rs)}_{\nu,J}(P_i)\,,
\eea
where the parenthesis in the integrand denotes the Euclidean inner product.
Computing the inner product of a diagram $\cA(P_i)$ and $\Psi_{\nu, J}^{(\rs)}(P_i)$, we can obtain the spectral function for the diagram $\cA(P_i)$\,.
After performing the $\nu$-integral by picking up poles in the spectral function, we can obtain the conformal block decomposition of the diagram $\cA(P_i)$\,.

%%%%%%%%%%%%%%%%%%%%%%
%%%%%%%%%%%%%%%%%%%%%%
\subsection{Contact Diagram}
\paragraph{}
Next we consider the conformal block decomposition of a contact diagram with $\phi^4$ interaction:
\bea
\cA^{\phi^4}(P_i) = \int_{\text{AdS}}dX ~ \prod_{i=1}^4 \frac{\cC_{\Delta_i}}{(-2 P_i\cdot X)^{\Delta_i}}\,.
\eea
The spectral integral for this diagram is given through the inversion formula: 
\bea \label{3: SR-contact}
\cA^{\phi^4}(P_i) = 
\sum_{J=0}^\infty \int_{-\infty}^\infty \frac{d\nu}{n_{\nu,J}} ~ 
\left(\cA^{\phi^4}, {\Psi}^{(\rs)}_{\nu,J}\right) \, 
\Psi^{(\rs)}_{\nu,J}(P_i)\,.
\eea
In the following, we will compute the inner product in the above integration.
After the computation, the poles in the spectral function tell us what kind operators are contained in the contact diagram.
Using the bulk representation of $\Psi_{\nu, J}^{(\rs)}(P_i)$ and gluing the bulk-to-boundary propagators through the AdS harmonic function,
the inner product is evaluated as the following bulk diagram:
\bea
\left(\cA^{\phi^4}_1,{\Psi}^{(\rs)}_{\nu,J}\right) 
&=&
\frac{1}{\pi^h}\left(\prod_{i=1}^4 \frac{\pi}{\alpha_i^2} \right)\frac{\pi}{\nu^2}
\frac{1}{\cB^{d-\Delta_1,d-\Delta_2}_{h-i\nu,J} \cB^{d-\Delta_3,d-\Delta_4}_{h+i\nu,J}}
\left[\frac{1}{J!(h-\frac{1}{2})_J}\right]^2~\int_{\text{AdS}_{d+1}} \frac{dXdX_LdX_R}{\text{vol}\left(SO(1,d+1)\right)}\nn\\[8pt]
&& 
\times
\Omega_{\alpha_1}(X,X_L)~(K_L\cdot \nabla_L)^J \Omega_{\alpha_2} (X,X_L)
\Omega_{\alpha_3}(X,X_R)~(K_R\cdot \nabla_R)^J \Omega_{\alpha_4} (X,X_R)\nn\\
&&
\times
\Omega_{\nu,J} (X_L,X_R;W_L,W_R)\,.
\eea 
\vspace{-1cm}
\begin{figure}[h]
\begin{center}
\includegraphics[height=5cm]{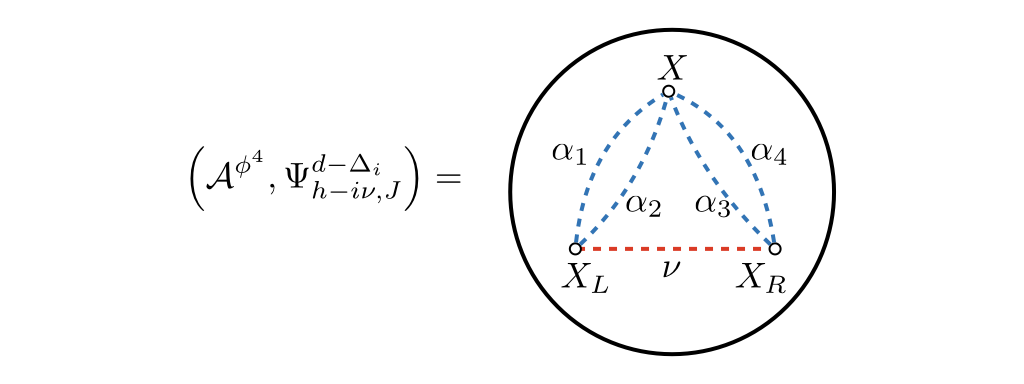}
\caption{\label{Fig:IP-contact}
The inner product $\left(\cA^{\phi^4}, \Psi^{(s)}_{\nu,J}\right)$ as a bulk diagram. }
\end{center}\end{figure}
%%%%%%%%%%%%%%%%%%%%%%
\paragraph{}
Next we will focus on the $X_L$ integration. 
This integration has the almost same structure as $\Xi^{\alpha_1,\alpha_2}_{\nu, J}$ function which is defined in \eqref{3: Def:Xi}
except for the derivatives at the bulk point $X$\,.
Due to the lack of derivatives within the interaction vertex for contact Witten digram, only $\Xi^{\alpha_1,\alpha_2}_{\nu, J}$ with $J=0$ can have non-vanishing value, 
and it is easily evaluated as:
\bea
\Xi^{\alpha_1,\alpha_2}_{\nu, 0}(X, X_R)&= &
\int dX_L~
\Omega_{\alpha_1}(X,X_L)\Omega_{\alpha_2} (X,X_L)\Omega_{\nu} (X_L,X_R)\nn\\
&=& F(\alpha_1,\alpha_2,\nu) \Omega_{\nu} (X,X_R)\,.
\eea
Now the $X_R$ integration is also easily computed using the previous formula, and we obtain the following result:
\bea
\left(\cA^{\phi^4},{\Psi}^{(\rs)}_{\nu,J}\right) 
&= &
\frac{1}{\pi^h}\left(\prod_{i=1}^4 \frac{\pi}{\alpha_i^2} \right)\frac{\pi}{\nu^2}
\frac{F(\alpha_1,\alpha_2,\nu)F(\alpha_3,\alpha_4,\nu)}{\cB^{d-\Delta_1,d-\Delta_2}_{h-i\nu,J} \cB^{d-\Delta_3,d-\Delta_4}_{h+i\nu,J}}
 \times \delta_{J,0}~\Omega_{\nu}(X,X) \int_{\text{AdS}} \frac{dX}{\text{vol}\left(SO(1,d+1)\right)}\,.\nn
\eea
The result is proportional to a Kronecker's delta $\delta_{J,0}$ according to the reason argued above.
Substituting this result into \eqref{3: SR-contact}, we obtain the spectral representation for the contact diagram:
\bea
\cA^{\phi^4}(P_i)
&=&
N^{\phi^4}\int_{-\infty}^\infty \frac{d\nu}{2 \pi}~
\omega^{\phi^4}_0(\nu)
~\Psi^{(\rs)}_{\nu,J}\,,\nn\\
\omega^{\phi^4}_0(\nu)&=& 
\Gamma\left(\frac{\Delta_{12}^+-h\pm i\nu}{2}\right)
\Gamma\left(\frac{\Delta_{34}^+-h\pm i\nu}{2}\right)
\frac{\Gamma\left(\frac{h+i\nu\pm\Delta_{12}^-}{2}\right)\Gamma\left(\frac{h-i\nu\pm\Delta_{34}^-}{2}\right)}{8~\Gamma(\pm i\nu)}\,.
\eea
Now the function omega is called the spectral function because this function is regarded as the integration kernel
and its pole structure determines the spectrum of operators in the conformal block decomposition.
It is obvious that the spectral function contains sets infinite series of double trace poles at $h+ i\nu=\Delta_{12}^+ + 2n$ and $h+ i\nu = \Del_{34}^+ + 2n$, $n=0,1,2, \dots$
The remaining poles are unphysical, and these are canceled with the zeros in the coefficient $\bc_{h-i\nu,0}^{(\rs)}$.
We can perform the $\nu$-integration picking up the relevant poles in the gamma functions and 
obtain the conformal block decomposition of the contact diagram.
This result is consistent with the fact that the contact diagram can be decomposed into conformal blocks of scalar double trace operators, 
and the coefficients we obtain from the residues match with the results obtained from recursive approach in \cite{Zhou:2018}.
%%%%%%%%%%%%%%%%%%%%
%%%%%%%%%%%%%%%%%%%%
\subsection*{Acknowledgement}
The authors would like to thank  Aninda Sinha and Chung-I Tan for very useful discussions.
The authors are also very grateful to Tatsuma Nishioka for commenting on the draft and collaborating on related projects.
This work of HYC was supported in part by Ministry of Science and Technology (MOST) 
through the grant 107-2112-M-002-008-, and Center for Theoretical Sciences, National Taiwan University.
The work of HK was supported by the Japan Society for the Promotion of Science (JSPS) and by the Supporting Program
for Interaction-based Initiative Team Studies (SPIRITS) from Kyoto University.
The authors are grateful to the organizers of New Frontiers in String Theory 2018, East Asian Joint Workshop on Fields and Strings 2018 and Indian Strings Meeting 2018 for the opportunities to present some of the work covered here, HYC is also grateful to Tokyo University, Hongo Campus and Kyoto University for the hospitalities when this work is being completed. 

\appendix
%%%%%%%%%%%%%%%%%%%%%%
\section{Useful Facts and Identities for Hypergeometric Functions}\label{Appendix:Hyper}
\paragraph{}
In this appendix, we list out the definitions and useful identities for the hypergeometric functions with one and multiple variables used in the main text, following  \cite{Schlosser} and \cite{Exton1}.
Starting with the series definitions of regularized Gauss Hypergeometric function:
\be\label{Def:t2F1}
{}_2 \tilde{F}_1  \left[
\begin{matrix}
~a,~b\\
~c
\end{matrix}; \rv
\right]  = \sum_{m=0}^{\infty} \frac{(a)_m (b)_m}{\Ga(c+m)} \rv^m, ~~|\rv|<1, ~~{\rm Re}(c-a-b)>0
\ee 
and it satisfies the following identities:
{\small{\be\label{hyper-id1}
{}_2\tilde{F}_1 \left[
\begin{matrix}
~a, ~b \\
~c
\end{matrix}; 1-\rv \right] = \frac{\pi}{\sin\pi (c-a-b)}\left[\frac{1}{\Ga(c-a)\Ga(c-b)} {}_2\tilde{F}_1 \left[
\begin{matrix}
~a, ~b \\
~a+b-c+1
\end{matrix};\rv \right]-\frac{\rv^{c-a-b}}{\Ga(a)\Ga(b)} {}_2\tilde{F}_1 \left[
\begin{matrix}
~c-a, ~c-b \\
~1-(a+b-c)
\end{matrix}; \rv \right] \right],
\ee}}
\be\label{hyper-id2}
{}_2\tilde{F}_1 \left[
\begin{matrix}
~a, ~b \\
~c
\end{matrix}; 1-\rv \right] = \frac{\pi}{\sin\pi (b-a)}\left[\frac{\rv^{-a}}{\Ga(b)\Ga(c-a)} {}_2\tilde{F}_1 \left[
\begin{matrix}
~a, ~c-b \\
~1+a-b
\end{matrix}; \frac{1}{\rv} \right]-\frac{\rv^{-b}}{\Ga(a)\Ga(c-b)} {}_2\tilde{F}_1 \left[
\begin{matrix}
~b, ~c-a \\
~1-a+b
\end{matrix}; \frac{1}{\rv} \right] \right].
\ee
We also consider Appell's hypergeometric function $\bF_4$, which can be defined as the following double infinite series:
\bea
 {\bf{F}}_4   \left[
\begin{matrix}
~a_1,~a_2\\
~c_1,~c_2
\end{matrix}; x, y
\right] &=& \sum_{m=0}^{\infty} \sum_{n=0}^{\infty} \frac{(a_1)_{m+n}(a_2)_{m+n}}{m! n! (c_1)_m (c_2)_n} x^m y^n, \quad |x|^{\frac{1}{2}} + |y|^{\frac{1}{2}} <1,\nn\\
&=&\Ga(c_1)\Ga(c_2)\sum_{m=0}^{\infty} \frac{(a_1)_m (a_2)_m}{m! \Ga(c_1+m)} x^m  {}_2 \tilde{F}_1  \left[
\begin{matrix}
~a_1+m,~a_2+m\\
~c_2
\end{matrix}; y
\right]\nn\\
&=&\frac{\Ga(c_1)\Ga(c_2)}{\Ga(a_1)\Ga(a_2)}\int^{i\infty}_{-i\infty}\frac{ds}{2\pi i} \frac{\Ga(a_1+s)\Ga(a_2+s)\Ga(-s) (-x)^{s}}{\Ga(c_1+s)} {}_2 \tilde{F}_1  \left[
\begin{matrix}
~a_1+s,~a_2+s\\
~c_2
\end{matrix}; y
\right].\nn\\
 \label{Def:F4}
 \eea
 The power series expression for $\bF_4$ \eqref{Def:F4} is in fact satisfies the following associated partial differential equations:
\bea
&&x(1-x)\frac{\partial^2 \bF_4}{\partial x^2}-y^2 \frac{\partial^2 \bF_4}{\partial y^2}-2xy \frac{\partial^2 \bF_4}{\partial x \partial y} +(c_1-(a_1+a_2+1)x) \frac{\partial \bF_4}{\partial x}-(a_1+a_2+1) y\frac{\partial \bF_4}{\partial y}-a_1a_2 \bF_4 = 0,\nn\\
\label{F4-eqn1}\\
&&y(1-y)\frac{\partial^2 \bF_4}{\partial y^2}-x^2 \frac{\partial^2 \bF_4}{\partial x^2}-2xy \frac{\partial^2 \bF_4}{\partial x \partial y} +(c_2-(a_1+a_2+1)x) \frac{\partial \bF_4}{\partial y}-(a_1+a_2+1) x\frac{\partial \bF_4}{\partial x}-a_1a_2 \bF_4 = 0,\nn
\label{F4-eqn2}\\
\eea
near the origin $(x, y) = (0, 0)$. These are analogous to Gauss's hypergeometric equation which admit different series solutions in different regions of $(\ru, \rv)$.
Using the integral representation \eqref{Def:F4} and the hypergeometric function identity \eqref{hyper-id2}, we can show that ${\bf{F}}_4$ obeys the following identity:
\bea\label{F4-Id1}
&&\bF_4 \left[
\begin{matrix}
~a_1,~a_2\\
~c_1,~c_2
\end{matrix}; x, y
\right] \nn\\
&&= \frac{\Ga(c_2)\Ga(a_2-a_1)}{\Ga(c_2-a_1)\Ga(a_2)} \frac{1}{(-y)^{a_1}} \bF_4 \left[
\begin{matrix}
~a_1,~a_1-c_2+1\\
~c_1,~a_1-a_2+1
\end{matrix}; \frac{x}{y}, \frac{1}{y}
\right]+\frac{\Ga(c_2)\Ga(a_1-a_2)}{\Ga(c_2-a_2)\Ga(a_1)} \frac{1}{(-y)^{a_2}} \bF_4 \left[
\begin{matrix}
~a_2,~a_2-c_2+1\\
~c_1,~a_2-a_1+1
\end{matrix}; \frac{x}{y}, \frac{1}{y}
\right],\nn\\
\eea
which is useful for evaluating the double commutator \eqref{double-comm-u}.
Finally $\bF_4$ also enjoys the highly non-trivial factorization and summation identity \cite{BC-Id}:
\bea\label{F4-Id2}
&&\bF_4 \left[
\begin{matrix}
~a_1,~a_2\\
~1+c_1,~1+c_2
\end{matrix}; x(1-y), y(1-x)
\right] \nn\\
&&=\sum_{m=0}^\infty \frac{(a_1)_m (a_2)_m (a_1+a_2-c_1-c_2-1)_m}{m! (1+c_1)_m (1+c_2)_m}  x^m y^m {}_2 F_1 \left[
\begin{matrix}
~a_1+m, ~a_2+m \\
~1+c_1+m
\end{matrix}; x \right] {}_2 F_1 \left[
\begin{matrix}
~a_1+m, ~a_2+m \\
~1+c_2+m
\end{matrix}; y \right]\nn\\
&&=\frac{\Ga(1+c_1)\Ga(1+c_2)}{\Ga(a_1)\Ga(a_2)}\sum_{m=0}^\infty \frac{\Ga(a_1+m) \Ga(a_2+m) (a_1+a_2-c_1-c_2-1)_m}{m! } \tilde{g}^{(m)}  \left[
\begin{matrix}
~a_1, ~a_2 \\
~1+c_1
\end{matrix}; x \right] 
\tilde{g}^{(m)}  \left[
\begin{matrix}
~a_1, ~a_2 \\
~1+c_2
\end{matrix}; y \right] 
\nn\\
\eea
where we have defined:
\be\label{Def:gm}
\tilde{g}^{(m)}  \left[
\begin{matrix}
~a, ~b \\
~c
\end{matrix}; z \right] = z^m   {}_2 \tilde{F}_1 \left[
\begin{matrix}
~a+m, ~b+m \\
~c+m
\end{matrix}; z \right].
\ee
In the main text, when expressing conformal blocks in terms of $\tilde{\bF}_4$, e. g. \eqref{Gblock-s}, we have $\ru = z\bz$ and $\rv = (1-z)(1-\bz)$, and $a_1+a_2-c_1-c_2-1 = h-1$, we can consider the symmetric combination:
\bea\label{Factorized-tF4}
&&\tilde {\bf{F}}_4   \left[
\begin{matrix}
~a_1,~a_2\\
~1+c_1,~1+c_2
\end{matrix}; \ru, \rv 
\right]  = \nn\\
&&\frac{\pi^2}{2\sin\pi c_1\sin\pi c_2} \sum_{m=0}^\infty \frac{\Ga(a_1+m)\Ga(a_2+m)}{m!} (h-1)_m \left\{ \tilde{g}^{(m)}  \left[
\begin{matrix}
~a_1, ~a_2 \\
~1+c_1
\end{matrix}; z \right] \tilde{g}^{(m)}  \left[
\begin{matrix}
~a_1, ~a_2 \\
~1+c_2
\end{matrix}; 1-\bz \right] +(z\leftrightarrow \bz) \right\}.\nn\\
\eea
In particular when $h=1$, we have
\be
{\bF}_4 \left[
\begin{matrix}
~a_1,~a_2\\
~1+c_1,~1+c_2
\end{matrix}; \ru, \rv
\right] =\frac{1}{2}\left\{{}_2{F}_1 \left[
\begin{matrix}
~a_1, ~a_2 \\
~1+c_1
\end{matrix}; z \right] {}_2 {F}_1 \left[
\begin{matrix}
~a_1, ~a_2 \\
~1+c_2
\end{matrix}; 1-\bz \right] + (z\leftrightarrow \bz)\right\}.\ee
This identity allows us to directly recover the $J=0$ conformal block in two dimensions.
%%%%%%%%%%%%%%%%%%%%%%%%%
%%%%%%%%%%%%%%%%%%%%%%%%%
\section{The Computation of $\Xi_{\nu, J}^{\alpha_1, \alpha_2}(X_1, X_2; W_1, W_2)$}\label{Appendix:Xi}
\paragraph{}
Here we will show the details of computation of $\Xi_{\nu, J}^{\alpha_1, \alpha_2}(X_1, X_2; W_1, W_2)$ introduced in \eqref{3: Def:Xi}, and the result is proportional to AdS harmonic function: 
\bea\label{3: Xi: omega}
\Xi^{\alpha_1,\alpha_2}_{\nu, J} (X_1,X_2;W_1,W_2) =F(\alpha_1,\alpha_2,\nu)\, \Omega_{\nu,J} (X_1,X_2;W_1,W_2)\,,
\eea
where the coefficient of proportionality $F(\alpha_1,\alpha_2,\nu)$ is given by: 
\bea
F(\alpha_1,\alpha_2,\nu) = 
\frac{J!\, \pi^h}{2^{J-1}\, \Gamma(h+J)}
\left(\prod_{i=1}^2\frac{\alpha_i^2}{\pi}\right)\,
\frac{\nu^2}{\pi}
\frac{1}{\cC_{h\pm i\nu,J}}
\cB^{\Delta_1,\Delta_2}_{h+i\nu;J}\,
\cB^{\bar{\Delta}_1,\bar{\Delta}_2}_{h-i\nu;J}\,,
\eea
where $\bar{\Del}_i =d -\Del_i$.
Because of the completeness of the AdS harmonic function, the function $\Xi_{\nu, J}^{\alpha_1, \alpha_2}$ which depends on two bulk points
can also be expanded in terms of them.
\paragraph{}
Basically $\Xi_{\nu, J}^{\alpha_1, \alpha_2}$ contains one bulk integral and three boundary integral which come from the definition of the AdS harmonic function.
In terms of AdS harmonic function, $\Xi_{\nu, J}^{\alpha_1, \alpha_2}$ is expanded as the following integration:
\bea\label{Def:Xi}
\Xi^{\alpha_1,\alpha_2}_{\nu, J} (X_1,X_2;W_1,W_2)&=&
\frac{\cN_0}{J! (h-\frac{1}{2})_J} \int_{\partial{\rm AdS}_{d+1}} dP_0dP_1dP_2\int_{{\rm AdS}_{d+1}} dY
\frac{1}{(-2X_1\cdot P_1)^{h+i\alpha_1}}\frac{1}{(-2Y\cdot P_1)^{h-i\alpha_1}}
\nn\\
& \times&
(W_1\cdot \nabla_1)^J\frac{1}{(-2X_1\cdot P_2)^{h+i\alpha_2}}
\, (K_Y\cdot \nabla_Y)^J\frac{1}{(-2Y\cdot P_2)^{h-i\alpha_2}}\, 
\nn\\
&\times&
\frac{1}{J!(h-1)_J}
\frac{(-2W_2\cdot C^\cD_0\cdot X_2)^J}{(-2X_2\cdot P_0)^{h+i\nu+J}}
\frac{(-2W_Y\cdot C^Z_0\cdot Y)^J}{(-2 Y\cdot P_0)^{h-i\nu+J}}\,.\\
\cN_0
&=&\frac{\alpha_1^2\, \alpha_2^2\, \nu^2}{\pi^3}~ \cC_{h\pm i\alpha_1} \cC_{h\pm i\alpha_2} \cC_{h\pm i\nu,J}\,,
\eea
where in $C^\cD_0$, $Z_0^A$ is replaced with the differential operator $\cD_{Z_0}^A$ in $C_0^{AB} = Z_0^A P_0^B -P_0^A Z_0^B$ to perform contractions.
Firstly we focus on the following bulk integral:
\bea
\cI_Y\equiv \frac{1}{J! (h-\frac{1}{2})_J}\int_{{\rm AdS}_{d+1}} dY~ \frac{1}{(-2 P_1\cdot Y)^{h-i \alpha_1}}(K_Y\cdot \nabla_Y)^J~
\frac{1}{(-2 P_2\cdot Y)^{h-i\alpha_2}}~\frac{(-2W_Y\cdot C^Z_0\cdot Y)^J}{(-2P_0\cdot Y)^{h-i\nu +J}}\,.\nn\\
\eea
This integral is a usual three-point Witten diagram with two scalars and one tensor, and was already evaluated in \cite{SpinningAdS}:
\bea
\cI_Y&=&
\cN_Y
~ 
\frac{(-2P_1\cdot C_0\cdot P_2)^J}{P_{01}^{\gamma^{-+-}} P_{02}^{\gamma^{+--}} P_{12}^{\gamma^{--+}}}\,,\\
\cN_Y&=& \frac{(-2)^J\pi^h \Gamma\left(\gamma^{--+}\right)
\Gamma\left(\gamma^{-+-}\right)
\Gamma\left(\gamma^{+--}\right)
\Gamma\left(\gamma^{---}\right)}
{2\, \Gamma(h-i\alpha_1)\Gamma(h-i\alpha_2)\Gamma(h-i\nu+J)}\,,\nn
\eea
where $\gamma^{--+},...$ depend on the relative signs among the spectral parameters and are defined as:
\bea
\gamma^{\sigma_1\sigma_2\sigma_0}\equiv \frac{1}{2}(h+J+i(\sigma_1\, \alpha_1+\sigma_2\, \alpha_2+\sigma_0\, \nu))\,.
\eea
Next we focus on the boundary $P_1$ integral:
\bea
\cI_{1}=\int_{\partial{\rm AdS}_{d+1}} dP_1 
\frac{1}{(-2 P_1 \cdot X_1)^{h+i\alpha_1}} 
\frac{(-2 P_1 \cdot C^Z_0 \cdot P_2)^J}{(-2 P_0\cdot P_1)^{\gamma^{-+-}} (-2 P_2\cdot P_1)^{\gamma^{--+}}}\,.
\eea
Using the generalized Symanzik formula which is given in \cite{CKK2}, and introducing a Mellin variable $t$, it can be evaluated as:
\bea
\cI_{1}&=&\cN_1 
\int^{i\infty}_{-i\infty} \frac{dt}{2\pi i}~
\mu(t)~
\frac{(2P_2\cdot C^Z_0 \cdot X_1)^J}
{(-2P_0\cdot X_1)^{\gamma^{++-}+t}(-2P_2\cdot X_1)^{\gamma^{+-+}+t}(-2P_0\cdot P_2 )^{-t-i\alpha_1}}\nn\\
\cN_1&=& \frac{\pi^h}{\Gamma(h+i\alpha_1)\Gamma(\gamma^{-+-})\Gamma(\gamma^{--+})}\,,
\qquad \mu(t)=\Gamma(-t)\Gamma(-i\alpha_1-t)\Gamma(\gamma^{++-}+t)\Gamma(\gamma^{+-+}+t)\,.\nn
\eea
Now the original bulk integration has the following form:
\bea
&&\Xi^{\alpha_1,\alpha_2}_{\nu, J} (X_1,X_2;W_1,W_2) \\
&&=
\cN_0\, \cN_Y\, \cN_1\, \int_{\partial{\rm AdS}_{d+1}} dP_0\, dP_2
\left[(W_1\cdot \nabla_1)^J \, 
\frac{1}{(-2X_1\cdot P_2)^{h+i\alpha_2}}\right]
\frac{(-2 W_2\cdot C^\cD_0\cdot X_2)^J}{(-2 X_2\cdot P_0)^{h+i\nu+J}}\nn\\
&&\times  \frac{1}{J!(h-1)_J}
\int \frac{dt}{2 \pi i} 
\mu(t) \frac{(2P_2\cdot C^Z_0\cdot X_1)^J}
{(-2P_0\cdot X_1)^{\gamma^{++-}+t}
(-2P_0\cdot P_2)^{\gamma^{+--}-i\alpha_1-t}
(-2P_2\cdot X_1)^{\gamma^{+-+}+t}}
\nn
\eea
The remaining $P_2$ integration is also evaluated by the Symanzik formula:
\bea
\cI_2&=&
2^J\, (h+i\alpha_2)_J
\int d P_2~  
\frac{(W_1\cdot P_2)^J (2X_1\cdot C_0\cdot P_2)^J}
{(-2 X_1\cdot P_2)^{h+J+\gamma^{+++}+t} (-2 P_0\cdot P_2)^{\gamma^{---}-t}}\\
&=&\cN_2(t)\frac{(-2 W_1\cdot C_0\cdot X_1)^J}{(-2P_0\cdot X_1)^{\gamma^{---}-t}}
\nn\\
\cN_2(t)&=&(-1)^J (h+i\alpha_2)_J
\frac{J!\, \pi^h~\Gamma(\gamma^{+++}+t)}{\Gamma(h+J+ \gamma^{+++}+t)} \,.\nn
\eea
Eventually, $\Xi_{\nu, J}^{\alpha_1, \alpha_2}$ becomes a boundary $P_0$ integral and a Mellin integration, however
the boundary integration is the same integral in the definition of the AdS harmonic function with spin $J$.
Thanks to this fact, the integral is replaced with a harmonic function, and $\Xi_{\nu, J}^{\alpha_1, \alpha_2}$ becomes
\be
\Xi^{\alpha_1,\alpha_2}_{\nu, J} (X_1,X_2;W_1,W_2) =\cN_0\, \cN_Y\, \cN_1\, 
\int^{i\infty}_{-i\infty} \frac{dt}{2\pi i} \mu(t)\,\cN_2(t)~ \frac{\pi}{\nu^2\cC_{h\pm i\nu,J}}\Omega_{\nu,J}(X_1,X_2;W_1,W_2)\,.
\ee
The remaining $t$ integration gives the following gamma functions through the Barnes's second formula:
\be
\int^{i\infty}_{-i\infty} \frac{dt}{2\pi i} \mu(t)~\frac{\Gamma(\gamma^{+++}+t)}{\Gamma(h+J+ \gamma^{+++}+t)}
=
\frac{\Gamma(\gamma^{+++})\Gamma(\gamma^{++-})\Gamma(\gamma^{+-+})\Gamma(\gamma^{-+-})\Gamma(\gamma^{--+})\Gamma(\gamma^{-++})}
{\Gamma(h+J)\Gamma(h+i\alpha_2+J)\Gamma(h+i\nu+J)}
\ee
Finally we can conclude that the boundary integration $\Xi_{\nu, J}^{\alpha_1, \alpha_2}$ is proportional to a AdS harmonic function 
and the coefficient is given as the following expression:
\bea
\Xi^{\alpha_1,\alpha_2}_{\nu, J} (X_1,X_2;W_1,W_2) &=&
F(\alpha_1,\alpha_2,\nu)~ \Omega_{\nu,J}(X_1,X_2;W_1,W_2)\nn\\
F(\alpha_1,\alpha_2,\nu) &= &
\frac{J!\, \pi^h}{2^{J-1}\, \Gamma(h+J)}
\left(\prod_{i=1}^2\frac{\alpha_i^2}{\pi}\right)\,
 \frac{1}{\cC_{h\pm i\nu,J}}
 \cB^{\Delta_1,\Delta_2}_{h+i\nu;J}\,
\cB^{\bar{\Delta}_1,\bar{\Delta}_2}_{h-i\nu;J}\,.
\eea
Note here the AdS function is a even function in $\nu$, which means it is invariant under $\nu \rightarrow -\nu$,
and the function $F(\alpha_1,\alpha_2,\alpha_3)$ is also even in each $\alpha_i$.
The result of this calculation is summarized in Fig. \ref{3: fig Comp-Xi} in the main text.

\bibliographystyle{junsrt}

\end{document}